\documentclass[journal]{IEEEtran}

\usepackage{amsmath}
\usepackage{array}
\usepackage{amssymb}
\usepackage{amsfonts}
\usepackage{graphicx}
\usepackage{epsfig}
\usepackage{subfigure}
\usepackage{psfrag}
\usepackage{cite}
\usepackage{latexsym}
\usepackage{url}
\usepackage{color}
\usepackage{multirow}
\usepackage{mathtools}
\usepackage{bm}
\usepackage{booktabs}
\usepackage{algorithm}
\usepackage{algpseudocode}
\usepackage{bm}

\usepackage{verbatim}
\usepackage{indentfirst}
\usepackage{hyperref}
\usepackage{multirow}
\usepackage{makecell}
\usepackage{bbm}

\usepackage{booktabs}
\usepackage{tabularx}
\usepackage{array}
\usepackage{makecell}

\graphicspath{{fig/}}

\PassOptionsToPackage{bookmarks={false}}{hyperref}

\IEEEoverridecommandlockouts

\allowdisplaybreaks[4]

\usepackage{multirow}
\usepackage{wasysym}

\newcommand{\fullcircle}{\CIRCLE}     
\newcommand{\emptycircle}{\Circle}    
\newcommand{\halfcircle}{\LEFTcircle} 

\begin{document}
\title{Sense Smarter, Think Better: A Survey on Edge Perception for Next-Generation Networks}
\author{
Zhonghao Lyu, Xiaowen Cao, Xianxin Song, Yuchen Li, Jiacheng Wang, Shuoyao Wang, Yuanhao Cui, \\ Weijie Yuan, 
Xianghao Yu, Guangxu Zhu, 
Hai Liu, Jie Xu, \emph{Fellow, IEEE}, \\ Derrick Wing Kwan Ng, \emph{Fellow, IEEE}, and Shuguang Cui, \emph{Fellow, IEEE}
\thanks{Z. Lyu is with the Department of Computer Science, The Hang Seng
University of Hong Kong, Hong Kong, and with the Department of Information Science and Engineering, KTH Royal Institute of Technology, Stockholm, Sweden (e-mail: lzhon@kth.se).}
\thanks{X. Cao is with the College of Electronics and Information Engineering, Shenzhen University, Shenzhen 518060, China, and also with Guangdong Provincial Key Laboratory of Future Networks of Intelligence, Shenzhen  518172, China (e-mail: caoxwen@szu.edu.cn). }
\thanks{X. Song and X. Yu are with the Department of Electrical Engineering, City University of Hong Kong, Hong Kong, China (e-mail: xianxin.song@cityu.edu.hk, alex.yu@cityu.edu.hk). }
\thanks{Y. Li is with Baidu Inc., Beijing, China, and the School of Computer Science, Shanghai Jiao Tong University, Shanghai, China (e-mail: yuchenli1230@gmail.com).}
\thanks{J. Wang is with the College of Computing and Data Science, Nanyang Technological University, Singapore 639798 (e-mail: jiacheng.wang@ntu.edu.sg). }
\thanks{S. Wang is with the College of Electronic and Information Engineering, Shenzhen University, Shenzhen 518066, China (e-mail: sywang@szu.edu.com) }
\thanks{Y. Cui is with the School of Information and Communication Engineering, Beijing University of Posts and Telecommunications, Beijing 100876, China (e-mails: yuanhao.cui@bupt.edu.cn). }
\thanks{W. Yuan is with the School of Automation and Intelligent Manufacturing, Southern University of Science and Technology, Shenzhen 518055, China (email: yuanwj@sustech.edu.cn).}
\thanks{G. Zhu is with the Shenzhen Research Institute of Big Data, Shenzhen,
China (e-mail: gxzhu@sribd.cn).}
\thanks{J. Xu and S. Cui are with the School of Science and Engineering, the Shenzhen Future Network of Intelligence Institute (FNii-Shenzhen), and the Guangdong Provincial Key Laboratory of Future Networks of Intelligence, The Chinese University of Hong Kong (Shenzhen), Guangdong 518172, China (e-mail: xujie@cuhk.edu.cn, shuguangcui@cuhk.edu.cn). }
\thanks{H. Liu is with the Department of Computer Science, The Hang Seng
University of Hong Kong, Hong Kong (e-mail: hliu@hsu.edu.hk).}
\thanks{D. W. K. Ng is with the School of Electrical Engineering and Telecommunications, University of New South Wales, Sydney, NSW 2052, Australia (e-mail: w.k.ng@unsw.edu.au).}
}

\maketitle

\begin{abstract}
Edge perception has emerged as a foundational capability for future wireless networks, enabling the network edge to proactively sense, interpret, and interact with the physical environment in a task-oriented and resource-aware manner. This survey provides a comprehensive and structured overview of edge perception. We first review representative sensing modalities and edge artificial intelligence (AI) techniques as the fundamental building blocks. We then examine their synergistic interactions. We systematically analyze how edge AI enhances sensing capabilities, encompassing both in-band and out-of-band modalities, as well as multi-modal sensor data fusion. Moreover, we discuss the role of task-driven sensing in facilitating edge AI, including integrated sensing-communication-computation designs, and active perception frameworks that dynamically adapt sensing strategies for downstream applications. Finally, we identify key challenges and open issues.
By consolidating fragmented research across sensing, communication, and edge AI, this survey provides forward-looking insights for the design and implementation of edge perception systems for sixth-generation (6G) networks.
\end{abstract}


\section{Introduction}

\subsection{Research Background}
Future wireless networks are expected not only to deliver high-throughput and ultra-reliable communication services, but also to enable real-time and intelligent interaction with the physical environment. Recent ITU-R reports on IMT-2030 highlight the deep integration of sensing and artificial intelligence (AI) with communications as fundamental capabilities of the 6G era, indicating that perception and intelligence can no longer remain peripheral functions layered on top of a rate-centric network, but must instead be natively embedded within next-generation wireless systems \cite{ITU-R}.

On the sensing side, modern platforms, including smartphones, drones, industrial robots, and intelligent buildings, are increasingly equipped with a rich set of densely deployed sensors, such as cameras, LiDAR, millimeter-wave radars, microphones, and environmental- and bio- sensors \cite{10330577}. These heterogeneous and multi-modal sensing devices enable pervasive, fine-grained perception of the physical environment, generating continuous, high-dimensional data streams with unprecedented spatial and temporal resolution. Such sensing capabilities significantly enhance situational awareness and decision-making, supporting a wide range of complex and mission-critical applications, including autonomous driving, collaborative manufacturing, immersive extended reality, and large-scale environmental monitoring \cite{9810792}.

In parallel, the rapid development of edge AI, enabled by the advancements of both AI and the increasing availability of high-performance edge computing infrastructures, has fundamentally reshaped the AI execution paradigm \cite{9052677}. Traditional cloud-centric pipelines, which offload raw sensory data to centralized data centers, suffer from inherent latency, bandwidth, and privacy limitations when processing continuous, high-volume sensing workloads. Emerging applications demand end-to-end reaction times on the order of milliseconds, together with stringent reliability and context-awareness requirements. To effectively satisfy these demands, AI computation is increasingly migrated toward network-edge entities, such as base stations, access points, roadside units, and even end devices, enabling low-latency, near-source inference and on-device or edge-assisted learning. Such edge-native intelligence significantly reduces response latency, alleviates backbone congestion, and strengthens data privacy and security guarantees \cite{zhu2023pushing}.

\begin{figure*}[t]
  \centering
  \includegraphics[width=1\textwidth]{ 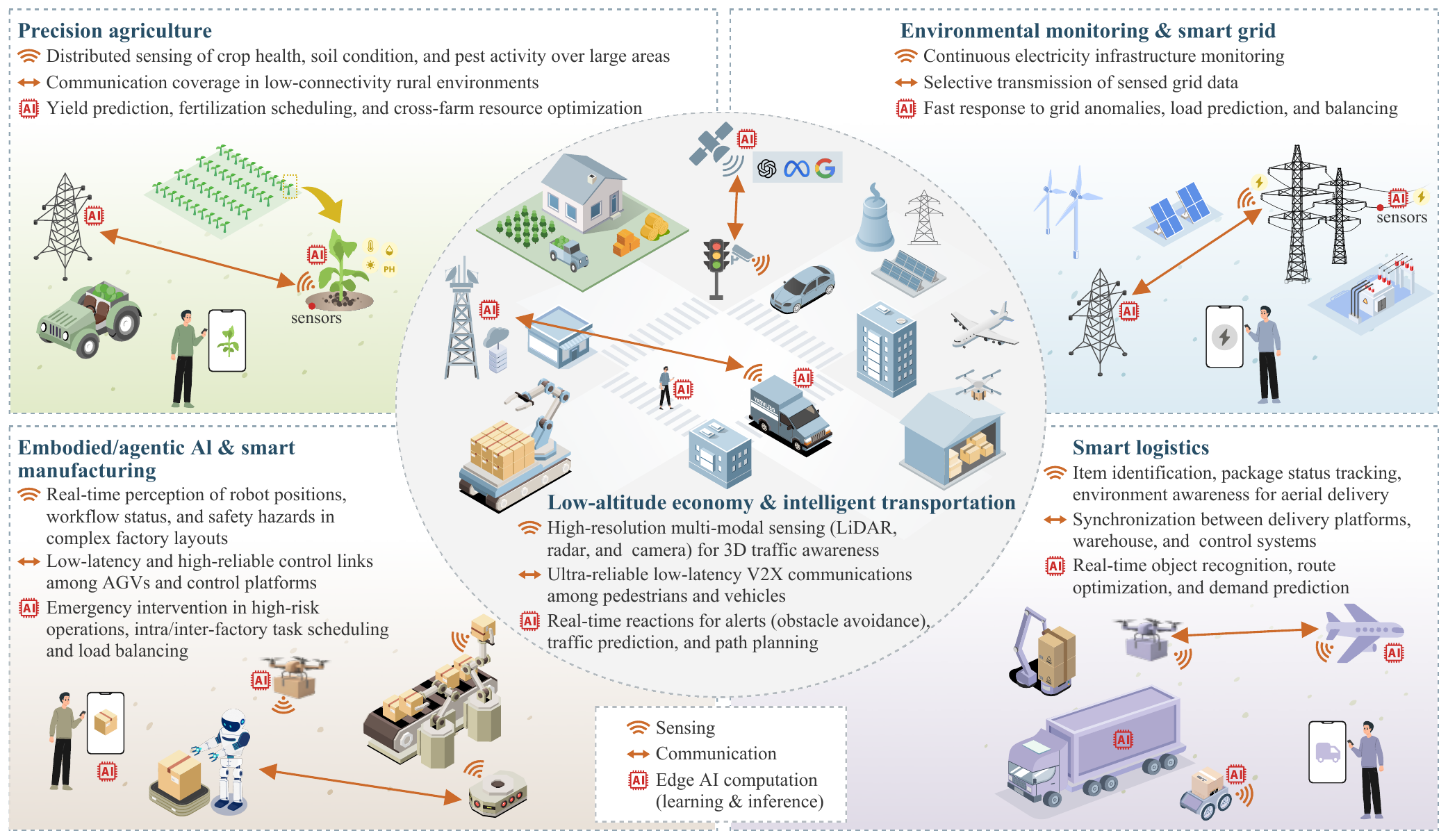}
  \caption{Representative application scenarios of edge perception, where sensing, communication, and edge AI are jointly integrated to support emerging services such as the low-altitude economy and embodied/agentic AI systems. Specifically, edge perception follows a closed-loop pipeline in which multi-modal sensing data are acquired and locally preprocessed at edge devices, transmitted to edge or cloud servers for further intelligent processing, and the results are fed back to adapt subsequent sensing, communication, and control actions.}
  \label{Fig:application}
\end{figure*}

The convergence of two technological trends, i.e., the ubiquitous deployment of high-precision multi-modal sensors and the shift of AI capabilities from centralized clouds to the network edge, inherently gives rise to a new edge-centric perception paradigm, referred to as edge perception. Instead of treating sensing as a passive data acquisition front end and AI as a remote, cloud-based post-processing stage, edge perception enables the network edge to actively sense, interpret, and react to environmental dynamics under stringent resource constraints and quality-of-service (QoS) requirements \cite{11098465}.

\subsection{Concepts and Applications of Edge Perception}

Edge perception aims to endow the network edge with the capability to proactively sense, interpret, and act upon the physical world in a task-oriented and resource-aware manner. To this end, an edge perception system tightly integrates heterogeneous sensing devices, wireless communication links, distributed edge computing resources, and AI models deployed across multiple edge nodes. Its processing pipeline spans the physical layer for signal acquisition and front-end processing, the network layer for data transport, coordination, and resource management, and the application layer for learning, inference, and control execution \cite{10908560}. Unlike traditional sensing systems that primarily aim to reconstruct high-fidelity signals or measurements, edge perception is driven by downstream tasks such as detection, tracking, recognition, and closed-loop control. Therefore, its performance is evaluated using task-level performance metrics (such as inference accuracy and control safety), rather than purely signal-oriented criteria.

 {A key feature of edge perception is the tight bidirectional coupling between sensing and edge AI. While AI models extract task-relevant representations from complex multi-modal signals, their effectiveness critically depends on the quality of sensing data. This interdependence enables AI to not only consume sensing information but also actively guide and adapt sensing behaviors, giving rise to a closed-loop and adaptive perception process.} Such a paradigm challenges conventional designs that treat sensing, communication, and AI as loosely coupled components and optimize them in isolation, which often struggle to achieve scalable and resource-efficient perception under increasing sensing density and AI workload complexity.
Throughout this paper, we adopt the following working definition:  {\emph{Edge perception refers to a task-oriented, end-to-end perception capability enabled by the coordinated design and operation of sensing, communication, and computation resources at the network edge, where sensing and edge AI are tightly integrated and mutually reinforcing.} }

Edge perception underpins a broad spectrum of emerging applications, such as the low-altitude economy (LAE) \cite{11318338}, large-scale environmental monitoring, and embodied and agentic AI systems, as shown in Fig.~\ref{Fig:application}. Across these domains, edge perception systems share a common sensing-communication-edge AI computation pipeline, while differing primarily in their dominant performance objectives, resource constraints, and operational environments. For example, safety-critical aerial and vehicular systems place stringent requirements on ultra-reliable and low-latency control, agricultural and environmental applications prioritize long-term robustness and energy efficiency under sparse connectivity, and industrial applications demand high-precision perception with strict privacy and security guarantees.

\begin{table*}[h]
\caption{Differences between existing surveys.}
\label{tab:survey_diff}
\centering
{\fontsize{6.5}{10}\selectfont
\setlength{\tabcolsep}{1pt}
\renewcommand{\arraystretch}{0.85}
\begin{tabular}{|c|l|c|c|c|c|c|c|c|}
\hline
 & \textbf{References} & \textbf{In-band sensing} & \textbf{Out-of-band sensing} & \textbf{Multi-modal fusion} &  \textbf{Edge learning} & \textbf{Edge inference} & \textbf{Integrated design} & \textbf{Closed-loop sensing-edge AI co-adaptation} \\
\hline

\multirow{9}{*}{\shortstack{Sensing-\\centric}}
 &  \cite{10.1145/3436729}        & \fullcircle & \emptycircle & \emptycircle &  \emptycircle & \emptycircle & \emptycircle & \emptycircle \\ \cline{2-9}
 &  \cite{9748867,10552143,10547188}  & WLAN   & \emptycircle & \emptycircle &  \emptycircle & \emptycircle & \emptycircle & \emptycircle \\ \cline{2-9}
 &    \cite{9900419}        & WLAN   & \emptycircle & \emptycircle & \emptycircle & \fullcircle  & \emptycircle & Edge AI for sensing \\ \cline{2-9}
 & \cite{9348922,10554983}       & mmWave      & \emptycircle & \emptycircle & \emptycircle & \emptycircle & \emptycircle & \emptycircle \\ \cline{2-9}
 & \cite{9455394,9106415}        & \emptycircle & Vision      & \emptycircle  & \emptycircle & \emptycircle & \emptycircle & \emptycircle \\ \cline{2-9}
 & \cite{10274950}            & \emptycircle & Vision      & \emptycircle   & \emptycircle & \fullcircle  & \emptycircle & Edge AI for sensing \\ \cline{2-9}
 & \cite{s23125406}             & \emptycircle & micro- and nano- sensors & \emptycircle  & \emptycircle & \fullcircle  & \emptycircle & \emptycircle \\ \cline{2-9}
 & \cite{s22155544}             & \emptycircle & IoT sensors & \emptycircle  & \emptycircle & \fullcircle  & \emptycircle & \emptycircle \\ \cline{2-9}
  & \cite{9810792}             & \emptycircle & IoT sensors & \emptycircle  & \fullcircle & \emptycircle  & \emptycircle & \bf{Co-adaptation} \\ \cline{2-9}
 & \cite{11127186}            & mmWave      & Camera      & \fullcircle    & \emptycircle & \emptycircle  & \emptycircle & Edge AI for sensing \\ 
\hline

\multirow{7}{*}{\shortstack{Edge-AI\\centric}}
 & \cite{8016573}             & \emptycircle & \emptycircle & \emptycircle & \emptycircle & \emptycircle & \emptycircle & \emptycircle \\ \cline{2-9}
 & \cite{9858872},\cite{10835069},\cite{9562559}   & \emptycircle & \emptycircle & \emptycircle &  \fullcircle  & \fullcircle  & ICC & \emptycircle \\ \cline{2-9}
 & \cite{10637271}            & \emptycircle & \emptycircle & \emptycircle  & \fullcircle  & \fullcircle  & ISCC & Task-oriented sensing assisted edge AI  \\ \cline{2-9}
 & \cite{10640100},\cite{10.1145/3581759}        & \emptycircle & \emptycircle & \emptycircle &  \fullcircle  & \fullcircle  & \emptycircle & \emptycircle \\ \cline{2-9}
 & \cite{10133894}, \cite{jsan11030047}        & \emptycircle & \emptycircle & \emptycircle &  \fullcircle  & \fullcircle  & ICC & \emptycircle \\ \cline{2-9}
 & \cite{10081195}           & \emptycircle & \emptycircle & \emptycircle  & \fullcircle  & \emptycircle & ICC & \emptycircle \\ \cline{2-9}
 & \cite{ren2023survey,9955525}      & \emptycircle & \emptycircle & \emptycircle & \emptycircle & \fullcircle & ICC & \emptycircle \\ 
\hline

\multirow{6}{*}{\shortstack{Integrated-\\design}}
 & \cite{9737357}            & \fullcircle & \emptycircle & \emptycircle  & \emptycircle & \fullcircle & ISAC & \emptycircle \\ \cline{2-9}
 & \cite{10418473}             & \fullcircle & \emptycircle & \emptycircle  & \emptycircle & \emptycircle & ISAC & \emptycircle \\ \cline{2-9}
 & \cite{10812728}            & \fullcircle & \emptycircle & \emptycircle  & \emptycircle & \fullcircle & ISCC & Task-oriented sensing assisted edge AI \\ \cline{2-9}
 & \cite{zhu2023pushing}             & \fullcircle & \emptycircle & \emptycircle  & \fullcircle  & \fullcircle  & ISCC & \emptycircle \\ \cline{2-9}
 & \cite{11098465}             & \fullcircle & \fullcircle & \halfcircle  & \fullcircle  & \fullcircle  & ISEA & \emptycircle \\ 
\hline

Ours & - & \fullcircle & \fullcircle & \fullcircle &  \fullcircle & \fullcircle & ISEA & \bf{Co-adaptation} \\
\hline

\multicolumn{9}{|l|}{\fullcircle: fully addressed \quad \halfcircle: partially addressed \quad \emptycircle: not mentioned} \\
\hline
\end{tabular}
}
\end{table*}

\subsection{Related Surveys}
In the literature, several survey papers have reviewed research topics closely related to edge perception from complementary perspectives, including sensing-centric surveys, edge-AI-centric surveys, and surveys on integrated sensing-communication-intelligence design. 

\subsubsection{Sensing-centric Surveys}
Existing sensing-centric surveys primarily examine how individual sensing modalities can be exploited for perception, typically treating sensing as an isolated or standalone processing pipeline. 
 {Existing sensing techniques can be broadly categorized into in-band and out-of-band sensing. While in-band sensing reuses communication waveforms and spectrum to perform sensing functions, out-of-band sensing relies on dedicated sensing modalities or signals operating outside the communication band.} In the literature of in-band sensing, \cite{10.1145/3436729} reviews the reuse of communication signals for sensing tasks such as gesture recognition and localization. Building on this general landscape, a number of surveys further specialize in specific wireless technologies. For example, \cite{9748867,10552143,10547188} survey commodity wireless local area network (WLAN)-based sensing, summarizing system architectures, signal representations, key performance indicators (KPIs), and representative applications enabled by channel state information (CSI), received signal strength indicator (RSSI) measurements, and emerging IEEE 802.11bf mechanisms. In addition, mmWave sensing surveys provide in-depth analyses of mmWave radar-based sensing systems, covering hardware platforms, signal processing pipelines, and representative applications in autonomous driving, smart homes, and industrial environments \cite{9348922,10554983}.

 {On the other hand, out-of-band sensing surveys have investigated LiDAR-, camera-, and micro-/nano-technology-enabled sensing for biomedical and environmental perception, focusing on underlying sensing principles, hardware architectures, and task-level algorithm design \cite{9455394,9106415,s23125406}.} In addition, collaborative sensing paradigms have been reviewed, including collaborative Internet of Things (IoT) sensing \cite{9810792} and radar-camera fusion for object detection and tracking \cite{11127186}, which summarize sensing models, data fusion strategies, quality metrics, and deployment considerations across heterogeneous devices. Despite their breadth, these surveys largely treat sensing as an isolated functional module, without considering how sensing  should be jointly organized, dynamically adapted, or coordinated under resource-constrained edge environments or explicitly aligned with downstream AI-driven tasks.

Several surveys further consider sensing in edge environments, emphasizing resource-efficient implementations on embedded platforms and lightweight model design. For instance, \cite{9900419} evaluates the feasibility of low-cost, real-time WiFi sensing on resource-constrained microcontrollers, while surveys on edge video analytics discuss system architectures and algorithmic techniques for executing deep vision models in proximity to cameras under stringent bandwidth and latency constraints \cite{10274950}. Context-aware edge sensing in wireless sensor networks has also been reviewed, highlighting the integration of local data processing and learning at the edge under resource limitations \cite{s22155544}. While these works take an important step by explicitly incorporating edge deployment constraints into sensing system design, they still largely decouple sensing and edge AI, without providing a systematic analysis of their tight bidirectional coupling and joint optimization.

\subsubsection{Edge AI-centric Surveys}
A large body of literature has surveyed edge artificial intelligence (AI) as a means to migrate AI functionalities from centralized clouds toward the network edge. As a foundation, a few surveys review mobile edge computing (MEC), covering system architectures, resource management, caching strategies, and task offloading mechanisms that support edge AI systems \cite{8016573,9858872}. Building upon this infrastructure, edge AI has been surveyed from multiple perspectives. For example, \cite{10637271} focuses on green edge AI, analyzing energy consumption and efficiency-oriented design principles, while \cite{10640100} reviews trustworthy edge AI, addressing security, reliability, transparency, sustainability, and their enabling mechanisms. Other surveys examine edge AI for large language models (LLMs) \cite{10835069} and lightweight machine learning (ML) \cite{10.1145/3581759}, summarizing architectures, compression techniques, and deployment strategies for resource-intensive or compact models at the network edge. In addition, domain-specific surveys further study edge AI in applications such as intelligent transportation systems \cite{10133894} and smart grids \cite{jsan11030047}.

Enabling edge AI crucially relies on how learning and inference are organized at the edge, and several surveys have focused on edge-native learning and inference paradigms. Federated and distributed learning in wireless networks have been systematically reviewed as mechanisms for collaborative training under resource constraints and network dynamics \cite{10081195,9562559}.
On the inference side, \cite{ren2023survey} surveys collaborative deep neural network (DNN) inference architectures that partition and distribute model execution across heterogeneous edge nodes to meet stringent latency and energy-efficiency requirements. Complementary tutorial works on semantic and task-oriented communications advocate aligning information exchange with downstream AI objectives, rather than optimizing bit-level reliability alone \cite{9955525}. Collectively, these surveys provide a comprehensive overview of edge AI architectures, algorithms, and system designs. However, they largely overlook how sensing functionalities are incorporated into edge AI systems, and rarely adopt a perception-centric viewpoint in which sensing and edge AI are jointly designed and optimized.

\subsubsection{Integrated-design Surveys}

Beyond sensing-centric and edge-AI-centric surveys, a third line of work investigates integrated frameworks that couple sensing, communication, and computation. Apart from the commonly adopted integrated communication and computation (ICC) design in edge-AI-centric surveys, a prominent example is integrated sensing and communication (ISAC), where radio access networks are envisioned to simultaneously support data transmission and environmental sensing. Recent ISAC surveys summarize application scenarios and analyze fundamental tradeoffs, physical-layer designs, and cross-layer resource management \cite{9737357, 10418473}.
Moving beyond dual-functional systems, several surveys have discussed integrated sensing, communication, and computation  (ISCC) designs \cite{zhu2023pushing,10812728}, in which sensing front ends, wireless links, and computing resources are orchestrated to optimize downstream task performance, together with signal design, resource management strategies, and open research challenges.

More recently, integrated sensing and edge AI (ISEA) has been proposed as a task-oriented paradigm that explicitly couples sensing with edge AI, jointly considering communication interfaces, AI computation, and sensing configurations for task-level optimization \cite{11098465}.
 {Different from ISAC, ISCC, or ISEA, edge perception is a perception-centric and end-to-end system capability that organizes heterogeneous sensing, data/feature exchange, edge AI execution, and downstream task feedback into a unified closed loop. While existing surveys on ISAC, ISCC, and ISEA \cite{9737357, 10418473,zhu2023pushing,10812728,11098465} study closely related enabling paradigms, they mainly adopt a wireless-system design perspective and emphasize the integration of sensing, communication, and computation over wireless links. In contrast, edge perception highlights a broader perception-centric abstraction, including a modality-level and fusion-oriented taxonomy across in-band and out-of-band sensing, as well as active perception or the closed-loop co-adaptation between sensing and edge AI.}

\begin{figure*}[h]
  \centering
  \includegraphics[width=0.98\textwidth]{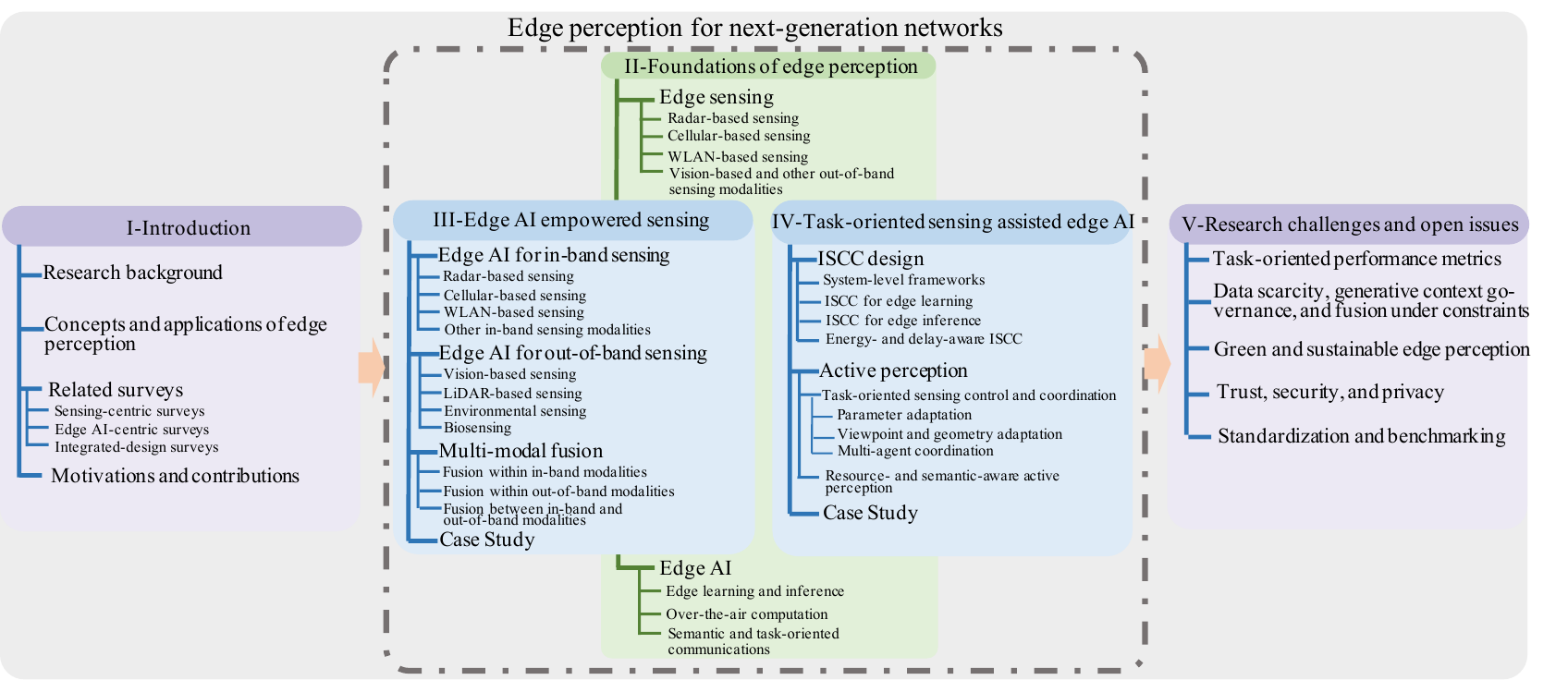}
  \caption{Structure of this paper, illustrating the closed-loop co-adaptation between edge sensing and edge AI through edge AI empowered sensing and task-oriented sensing-assisted edge AI.}
  \label{Fig:structure}
\end{figure*}

\subsection{Motivations and Contributions}
The above review of relevant surveys reveals that the existing literature remains fragmented when viewed from an edge-perception-centric perspective. Most prior surveys focus on individual sensing modalities, edge AI architectures, or narrowly scoped task-specific integrated designs, rather than treating edge perception as a unified, end-to-end system capability. Consequently, the intrinsic coupling and co-evolution between sensing and edge AI have not yet been systematically analyzed or articulated.

Motivated by this gap, rather than classifying techniques in parallel, this paper adopts a perception-centric viewpoint that focuses on how sensing and edge AI jointly enable task-oriented  perception at the network edge. In particular, we highlight the closed-loop interactions among sensing, communication, and intelligence that are critical for deployable edge perception systems. To this end, we systematically examine edge perception along three tightly coupled dimensions: edge sensing architectures and algorithms, edge AI foundations and models, and their bidirectional interplay, as shown in Fig. \ref{Fig:structure}. Together, these dimensions constitute a unified analytical framework for the convergence of perception and intelligence at the 6G edge. Accordingly, the main contributions of this survey are summarized as follows.
\begin{itemize}
    \item First, we formalize the concept of edge perception and establish a unified taxonomy, identifying edge sensing and edge AI as two fundamental and tightly coupled building blocks, and clarifying their respective roles, interfaces, and interactions across heterogeneous edge environments.
    \item Second, building upon this taxonomy, we provide a comprehensive survey of the bidirectional interplay between edge sensing and edge AI. We review how edge AI enhances sensing performance across both in-band and out-of-band modalities, and discuss multi-modal sensing architectures and data-fusion mechanisms. Moreover, we examine task-oriented sensing-assisted edge AI, including integrated ISCC designs and active perception mechanisms under practical and resource-constrained edge environments.
    \item Finally, we identify key research challenges and outline future research directions toward deployable edge perception systems, spanning task-oriented performance metrics, data and generative-content governance, system-level orchestration and lightweight model design, green and sustainable operation, and trust, security, and standardization issues at the network edge.
\end{itemize}

\begin{figure*}[h]
    \centering 
	\subfigure[Cellular-based sensing.]{\includegraphics[width=.32\textwidth]{ 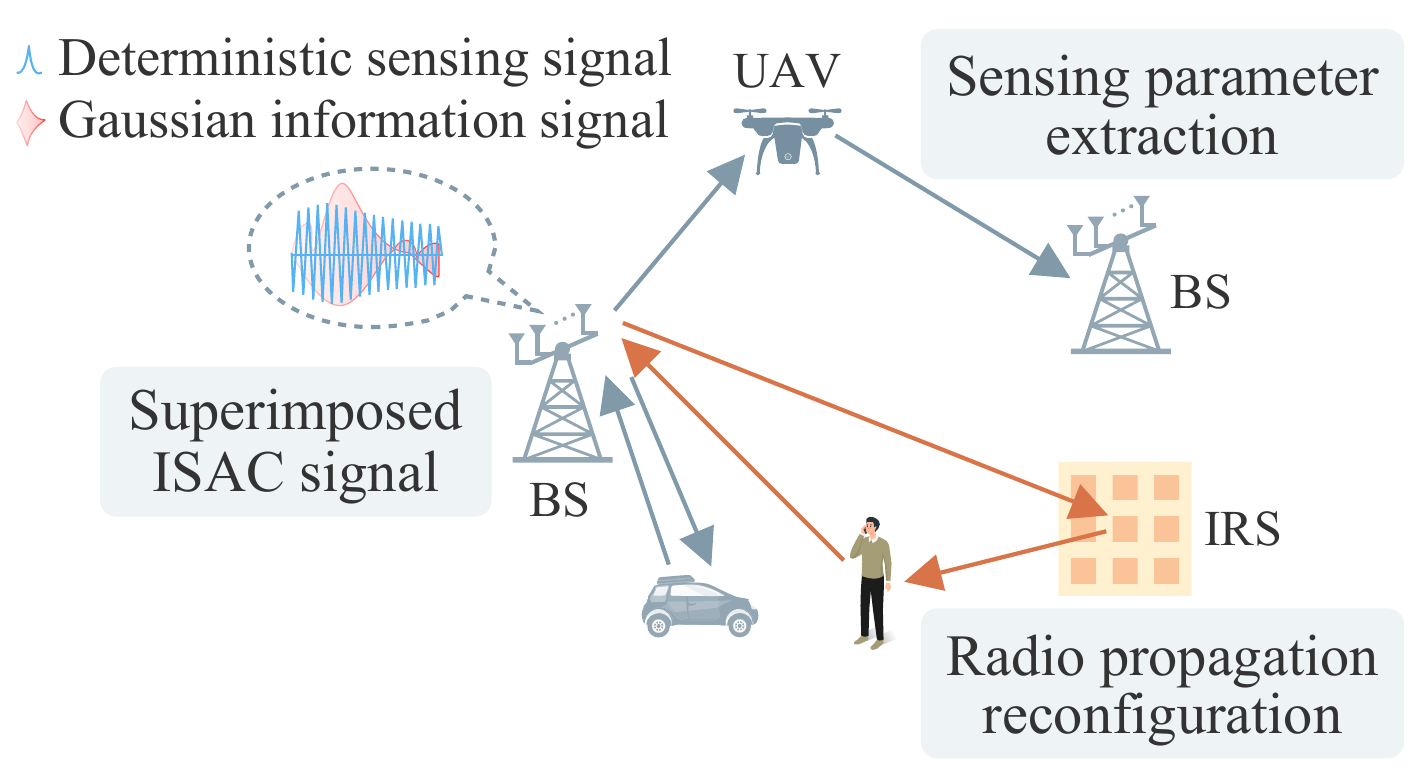}
    \label{Fig:Edgesensing_cellular}}
    \subfigure[WLAN-based sensing.]{\includegraphics[width=.32\textwidth]{ 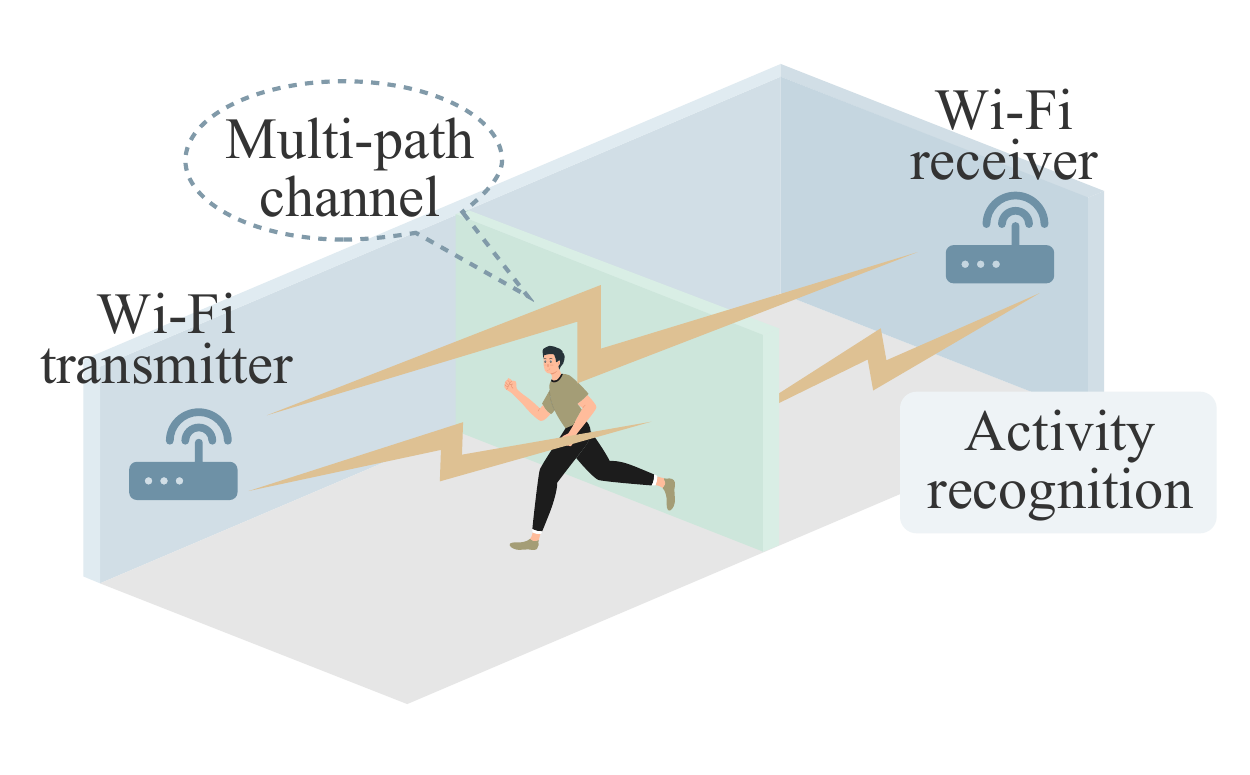}
    \label{Fig:Edgesensing_wlan}}
    \subfigure[Vision-based sensing.]{\includegraphics[width=.32\textwidth]{ 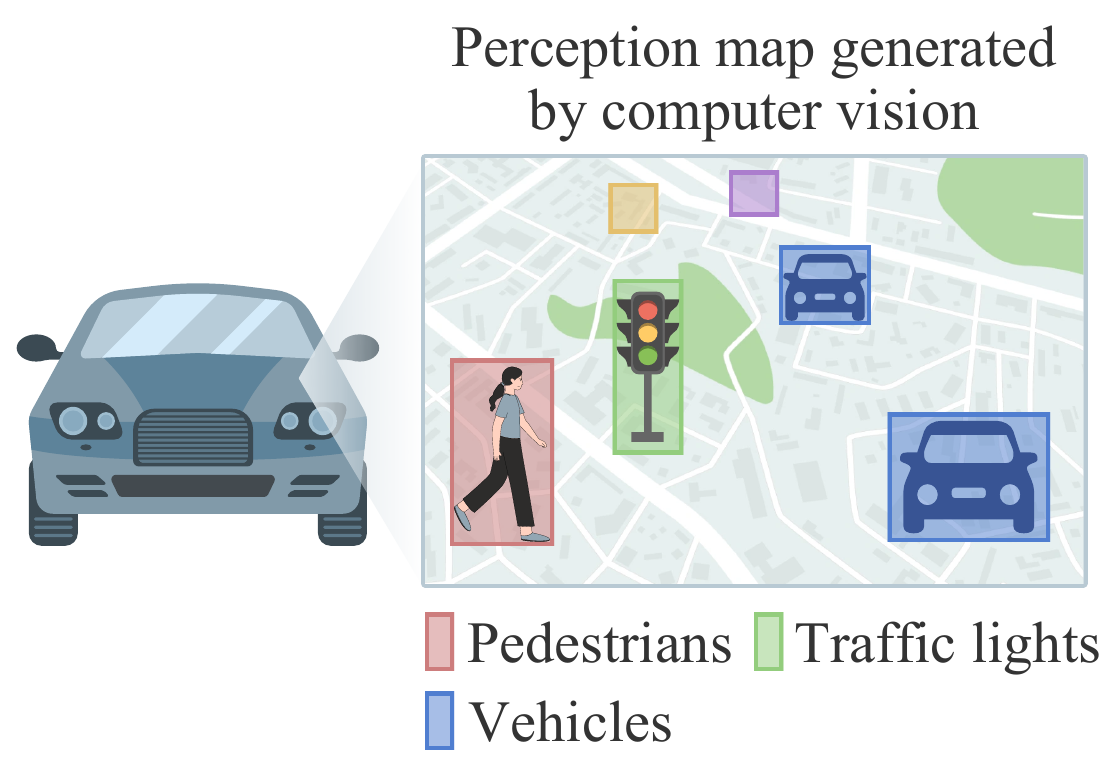}
    \label{Fig:Edgesensing_vision}}
    \caption{ {Representative types of edge sensing, including (a) cellular-based in-band sensing leveraging communication waveforms, (b) WLAN-based in-band sensing exploiting multipath propagation, and (c) vision-based out-of-band sensing relying on visual data for environmental perception.}}
\end{figure*}

\section{Foundations of Edge Perception}
Edge perception relies on two tightly coupled elements: 1) edge sensing, which captures information from the physical environment, and 2) edge AI, which processes sensed data into task-relevant knowledge and decisions under strict resource constraints.
This section reviews the foundations of edge perception from these two complementary perspectives.

\subsection{Edge Sensing}
We next review representative edge sensing systems focusing on architectures and receiver processing. Edge sensing is generally classified into in-band and out-of-band sensing, which reuses communication waveforms, and relies on dedicated sensors, respectively. For instance, radar-, cellular-, and WLAN-based sensing are typical in-band approaches, while vision-based sensing represents an out-of-band modality.

\subsubsection{Radar-based Sensing}
Traditional radar systems rely on transmitter-receiver pairs to extract sensing information by actively emitting probing waveforms and analyzing the resulting echo signals reflected from targets \cite{steven1993fundamentals,steven1993fundamentals_estimation}. 
These systems traditionally utilize dedicated  waveforms, such as pulse burst waveforms, frequency-modulated waveforms, and phase-coded waveforms as sensing signals. The sensing receivers extract key target parameters, including angle, velocity, and range, by analyzing the spatial spectrum, Doppler shifts, and propagation-induced time delays of the received echo signals. The perfectly known realizations and higher auto-correlation of sensing signals permit matched filtering and coherent integration at the sensing receiver, achieving superior performance. In radar sensing systems, parameter estimation and target detection are two representative sensing tasks. For parameter estimation, the Crame\'r-Rao bound (CRB) provides a fundamental limit on the parameter estimation error of any unbiased estimator \cite{steven1993fundamentals_estimation}, which is widely used to evaluate the sensing performance of practical estimators and facilitate the transmit signal design. The sensing performance of several established parameter estimation algorithms, including  maximum likelihood estimation (MLE), multiple signal classification (MUSIC), and estimation of signal parameters via rotational invariance techniques (ESPRIT), is close to or equal to the theoretical bounds \cite{schmidt1986multiple,32276}. For target detection, the probabilities of detection and false alarm are two key performance metrics, corresponding to the events in which the receiver correctly declares the presence of a target or incorrectly declares a target when none is present, respectively \cite{steven1993fundamentals}. Neyman-Pearson (NP)-based detectors and the generalized likelihood ratio test (GLRT) are two widely adopted detection frameworks, which aim to maximize the detection probability subject to a prescribed false-alarm probability constraint.

\subsubsection{Cellular-based Sensing}
Recently, the convergence of radar and communication systems in both hardware architectures and operating frequency bands, such as multi-antenna and millimeter-wave technologies, has enabled the seamless integration of sensing functionalities into wireless communication networks \cite{10188491,10547188,11103477}. 
As illustrated in Fig. \ref{Fig:Edgesensing_cellular}, the widespread deployment of communication networks provides a strong foundation for extending sensing coverage, especially with the advent of intelligent reflecting surfaces (IRS) and extremely large-scale MIMO (XL-MIMO) \cite{10500425,song2025overview,10138058,11216397}. Leveraging existing communication infrastructures for dual-purpose sensing can improve the spectrum utilization and energy efficiency, and is widely regarded as a key use case for future 6G networks \cite{zhi2025near}. Moreover, native connectivity among network nodes enables large-scale collaborative sensing through efficient data exchange, allowing spatially distributed nodes to jointly perform sensing, information fusion, and coordinated perception.  {This networked sensing capability fundamentally distinguishes cellular-based sensing from conventional standalone radar systems \cite{10879807}, as shown by the networking-based ISAC hardware testbed in \cite{JiComMag2023}, where two cooperative 5G NR ISAC systems reduce the positioning mean squared error (MSE) by 61\% over a single system while sustaining 2.8 Gbps throughput.}

\begin{table*}[h]
\centering
\caption{Pros and cons of representative edge sensing modalities.}
\label{Tab:ProsConsModalities}
\setlength{\tabcolsep}{0pt}
\renewcommand{\arraystretch}{1.4}
\begin{tabular}{
>{\raggedright\arraybackslash}p{0.11\linewidth}
>{\raggedright\arraybackslash}p{0.11\linewidth}
>{\raggedright\arraybackslash}p{0.4\linewidth}
>{\raggedright\arraybackslash}p{0.36\linewidth}}
\hline
\textbf{Category} & \textbf{Modality} & \textbf{Pros} & \textbf{Cons} \\
\hline

\multirow{6}{*}{\textbf{In-band}}
& Cellular
& Wide-area coverage ($>$100~{\rm m}), synchronized infrastructure, native network integration \cite{10188491,11103477,10500425,zhi2025near,10879807}.
& Waveform/resource coupling, multipath- and topology-dependent, limited semantic richness. \\
& WLAN
& Ubiquitous indoors (10-15~{\rm m}), fine-grained CSI for device-free sensing, low deployment cost \cite{5290370,9506934,8360860,8514811}.
& Device/environment heterogeneity, calibration/access issues, limited range/outdoor robustness \cite{9900419}. \\
& Radar
& Range-Doppler-angle, works in darkness, strong motion sensitivity \cite{steven1993fundamentals,steven1993fundamentals_estimation,richards2005fundamentals}.
& Clutter/interference, weaker semantics than vision. \\
\hline

\multirow{4}{*}{\textbf{Out-of-band}}
& Vision 
& Rich semantics, high spatial resolution, mature models/datasets \cite{10.1145/3495243.3517016,10540267,8567661,9769868}.
& Bandwidth/compute heavy (e.g., 10Mbps for 1080p videos), illumination/occlusion sensitivity, privacy concerns. \\
& LiDAR
& Accurate depth/3D geometry, mapping/localization friendly, illumination-invariant \cite{10.1145/3539491.3539591,8949733,10815971,10274112}.
& Weather degradation, higher cost, sparse/irregular point clouds \cite{9455394}. \\
& Environmental
& Low bandwidth/energy, privacy-friendly, ambient context cues \cite{s22155544}.
& Noisy and location-dependent, low spatial/semantic resolution. \\
& Biosensing
& Direct physiological states, personalization, continuous wearables \cite{bios15070410}.
& Motion artifacts, user robustness, strict privacy/security needs \cite{bios15070410}. \\
\hline
\end{tabular}
\end{table*}

Considering the intrinsic randomness of communication signals as well as the deterministic nature of dedicated sensing waveforms, prior research has explored transmit beamforming design and fundamental sensing performance limits in both monostatic  \cite{10251151,9916163,10147248,10596930,11087656,xie2024sensing} and bistatic \cite{song2025crb,song2025detection} sensing scenarios. 
In monostatic systems, the sensing transmitter and receiver are co-located. It is reasonable to assume that the sensing receiver has perfect knowledge of the transmitted signal, including its specific sequence. Existing works have followed two main approaches under this assumption. One line of works \cite{10251151,9916163} assumed that the sensing time is sufficiently long, so that the random signal realization of each snapshot approximates its statistical value. In this case, traditional performance metrics for deterministic signals are directly applied. Another line of works \cite{10147248,10596930,11087656,xie2024sensing} considered limited sensing intervals and employed the expectation of the performance metric over all possible signal realizations as the effective measure of sensing performance. Specifically, these works adopted the Bayesian CRB\cite{10147248}, ergodic linear MSE\cite{10596930}, expectation of the integrated sidelobe level (EISL)\cite{11087656}, and sensing mutual information (SMI) \cite{xie2024sensing} as sensing performance metrics, respectively. These works \cite{10147248,10596930,11087656,xie2024sensing} demonstrated that Gaussian information-bearing signals generally degrade sensing performance, and that the upper bound of sensing performance is achieved with deterministic sensing waveforms.


In bistatic sensing systems, the transmitter and receiver are separated. To perform coherent signal processing, the receiver must obtain the exact random signal sequences from the transmitter. This may create security issues and add significant communication overhead. A more promising approach is to bypass the transmission of specific signal realizations and use only their statistical properties for target sensing. The works in \cite{song2025crb,song2025detection} analyzed the parameter estimation CRB \cite{song2025crb} and detection probabilities \cite{song2025detection} with both deterministic and Gaussian information signals, respectively. These works \cite{song2025crb,song2025detection} showed that the sensing receiver can exploit the statistical properties of communication signals to improve sensing performance, even without knowledge of their specific realizations.


\subsubsection{WLAN-based Sensing}
WLAN-based sensing systems leverage existing WiFi hardware platforms and communication signals for environmental perception tasks, such as gesture recognition, fall detection, and environment imaging\cite{9941042,10188491} (as shown in Fig. \ref{Fig:Edgesensing_wlan}).  
WLAN-based sensing has the following advantages. First, since current WiFi standards natively support sensing-related functionalities,  no additional sensing hardware is required. Second, the global ubiquity of WiFi networks  implies that reusing WiFi signals for  sensing naturally complies with regional spectrum regulations, thereby facilitating large-scale and practical deployment. Third, WLAN-based sensing is a device-free sensing paradigm, which does not require  placing any devices on the sensing targets.

Generally, the presence of numerous static scatterers in WiFi scenarios creates complex multipath propagation and may block the line-of-sight (LoS)  path. This makes it difficult to directly obtain target parameters  by directly analyzing the spatial spectrum of received echo signals. Thus, WLAN-based sensing typically senses the environment by measuring some key signal characteristics, such as RSSI and CSI. In particular, the RSSI denotes the average received signal power at the sensing receiver, and has been used for coarse-grained localization and imaging tasks  \cite{5290370,9506934}. However, the received signal power is highly sensitive to multipath propagation characteristics and environmental dynamics. As a result, simplified  channel models cannot accurately capture the complex relationship between sensing targets and received signal power, which ultimately degrades overall performance.

To address this shortcoming, some previous works \cite{8360860,8514811} have performed environmental sensing by analyzing the CSI at the receiver. In contrast to the scalar power measurement of RSSI, CSI constitutes a finer-grained frequency-domain characterization of the channel, capturing more information such as complex amplitude and time delay over various subcarriers. To obtain  human-related parameters from the obtained CSI information, learning-based methods are  promising solutions. These approaches generally consist of two sequential stages: a training phase, in which models learn representative CSI characteristics across diverse scenarios, and an inference phase, where sensing results are generated from newly observed CSI measurements.

\begin{figure*}[h]
    \centering 
	\subfigure[]{\includegraphics[width=.32\textwidth]{ 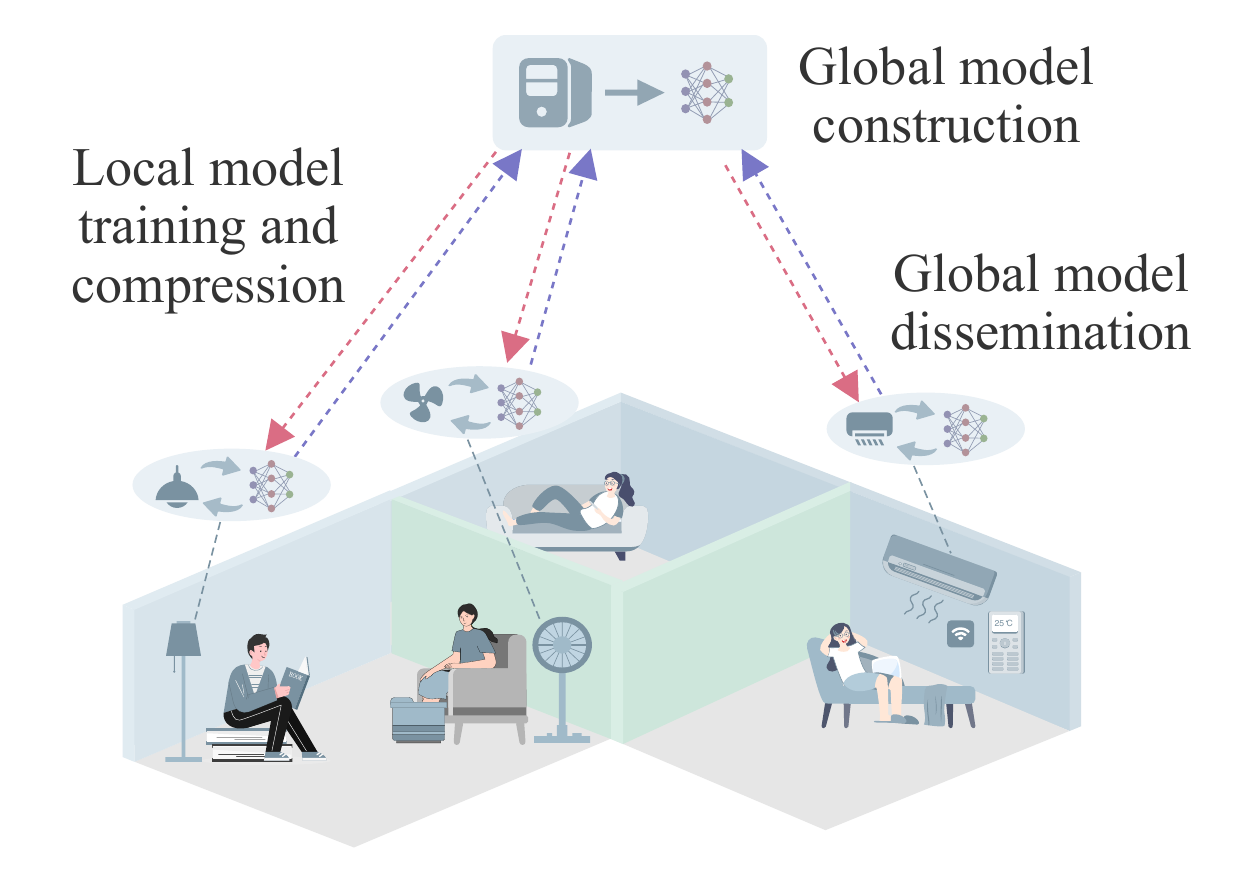}}
    \subfigure[]{\includegraphics[width=.32\textwidth]{ 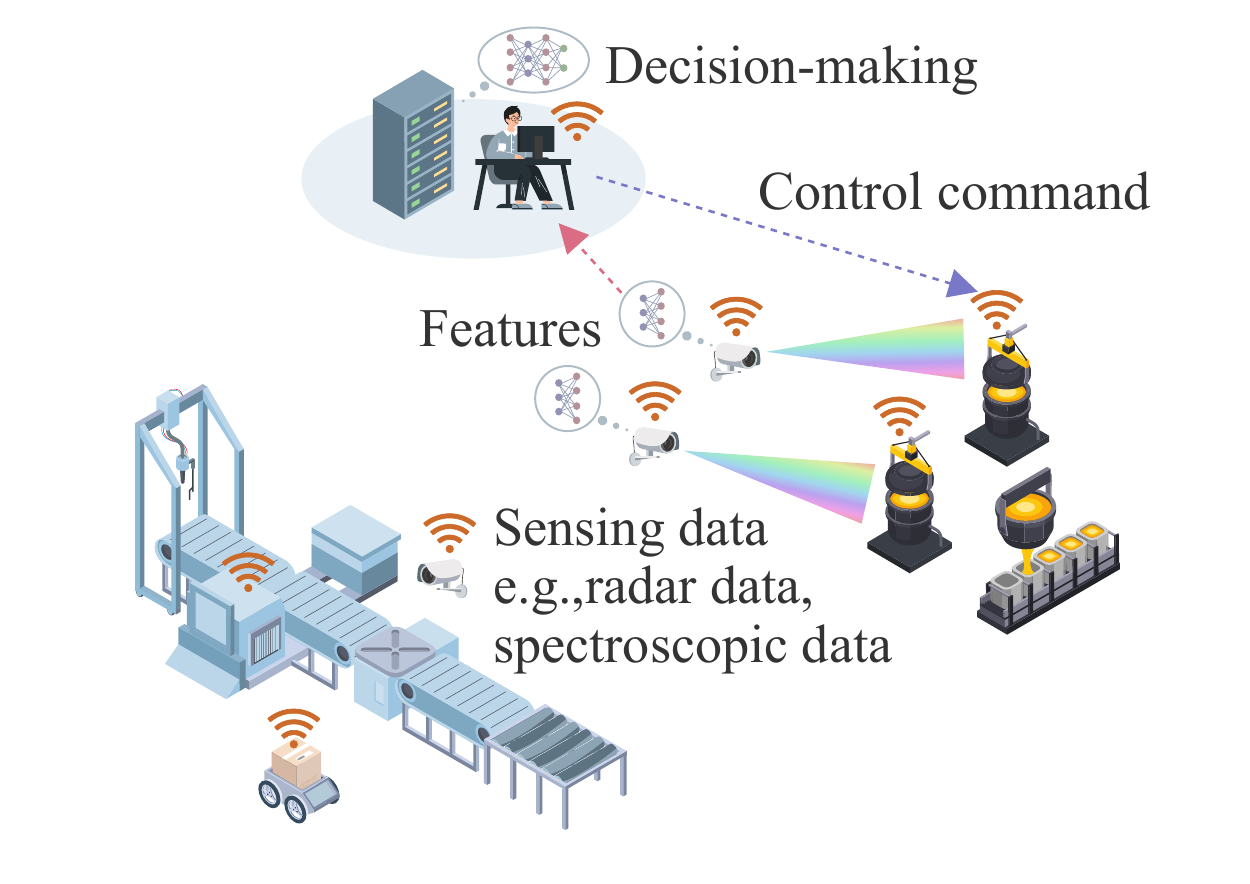}}
    \subfigure[]{\includegraphics[width=.32\textwidth]{ 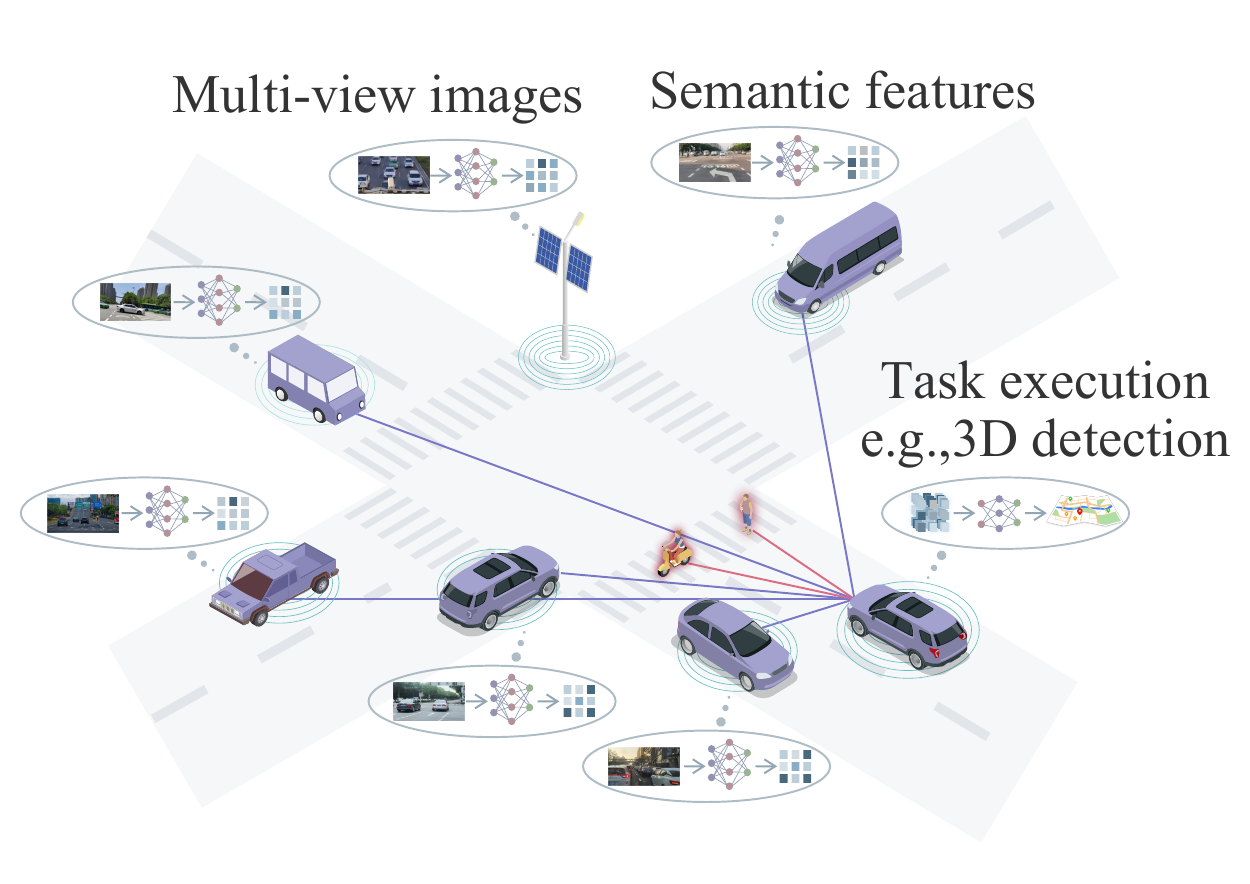}}
    \caption{ {Representative application scenarios of edge AI, illustrating (a) distributed edge learning for smart healthcare via wireless FL, (b) low-latency edge inference for industrial IoT, and (c) semantic/task-oriented communications for cooperative perception and decision-making in autonomous driving.}}
    \label{Fig:EdgeAI}
\end{figure*}

\subsubsection{Vision-based and Other Out-of-band Sensing Modalities}
Vision-based sensing is a key out-of-band edge perception modality, as it acquires environmental awareness by directly processing visual observations rather than repurposing communication waveforms, as shown in Fig. \ref{Fig:Edgesensing_vision}. In practice, cameras provide continuous, high-resolution, and semantically rich visual streams, enabling a broad spectrum of perception tasks, such as multi-object tracking and scene understanding \cite{6547194}. 

Beyond cameras, several other out-of-band modalities are increasingly integrated into edge perception systems. 
First, LiDAR provides accurate 3D geometry by measuring time-of-flight or phase shifts of laser pulses, yielding point clouds that are highly informative for depth estimation and 3D mapping \cite{haidar2024readyrealtimelidarsemantic}. 
Second, environmental sensing leverages low-cost sensors to capture ambient environmental context with low bandwidth \cite{11125738}. Third, biosensing modalities  offer direct and fine-grained measurements of physiological states that are difficult to infer reliably from external observations, thereby enabling personalized and health-aware edge perception \cite{bios15070410}. Table \ref{Tab:ProsConsModalities} summarizes representative in-band and out-of-band sensing modalities, highlighting their respective strengths and inherent limitations. In this context, multi-modal sensing and fusion are highly beneficial to enhance robustness, resolve ambiguities in challenging scenarios, and flexibly trade resource costs against task-level performance.

{\bf Lesson Learned:} The sensing techniques reviewed in this section are mainly rooted in classical signal processing methods, which play a fundamental role in extracting environmental information at the network edge. More recently, with their strong representation/reasoning capabilities, AI-based approaches have the potential to further enhance sensing performance. However, deploying AI models in edge environments is inherently constrained by limited resources, which necessitates careful adaptation of model architectures and inference pipelines. 
Moreover, edge sensing should not aim to indiscriminately acquire all available information. Given the tight coupling between sensing cost and resource consumption at the edge, effective sensing must be guided by downstream task objectives. Rather than maximizing generic sensing fidelity, sensing configurations should be adaptively tailored for task-relevant information that directly supports learning, inference, or decision-making. This task-oriented and adaptive sensing principle constitutes a key motivation for the closed-loop sensing-AI co-adaptation frameworks.

\subsection{Edge AI}

With advances in edge computing, AI functions are increasingly deployed at network-edge nodes. 
By performing training and inference close to data sources, edge AI reduces latency and improves data privacy, making it well suited for real-time edge perception applications \cite{zhu2023pushing}, as shown in Fig.~\ref{Fig:EdgeAI}. 

\subsubsection{ {Edge Learning and Inference}}
 {From a processing perspective, edge AI can be discussed across three main stages, namely pre-training, fine-tuning, and inference. Pre-training is typically conducted in the cloud due to its substantial data and computation demands. Edge learning more often takes the form of task-specific fine-tuning \cite{10791415}, distributed updating, and continual adaptation, which is particularly important for edge perception under dynamic and heterogeneous environments. Federated learning \cite{FedAvg} remains an important paradigm for privacy-preserving collaborative model training over distributed devices, as it can jointly exploit distributed sensing data and computation resources. Meanwhile, recent edge learning has moved beyond full-model updating toward parameter-efficient adaptation, where only a small subset of parameters is tuned. Representative examples include prompt-based federated adaptation and low-rank adaptation (LoRA)-based sparse updating for foundation models, which substantially reduce training overhead while retaining strong task adaptation capability \cite{10210127}. In addition, knowledge distillation and pruning-/quantization-aware training provide effective means to compress cloud-side knowledge into lightweight edge-deployable models, which is particularly beneficial for capacity-limited edge devices \cite{9964434,11270936}.}

 {Edge inference uses a well-trained model to make predictions or decisions on unseen data \cite{Shuvo23IEEE}. Modern edge inference has evolved beyond single-device execution and straightforward task offloading. A representative paradigm is split or collaborative inference, where model execution is partitioned across devices, edge servers, and clouds for joint execution. This improves the utilization of heterogeneous computation resources while reducing end-to-end latency and privacy exposure \cite{ren2023survey}. To further accelerate inference under stringent edge constraints, early-exit mechanisms enable adaptive computation-depth control according to the difficulty of each input sample \cite{9769868}. In addition, model cascading \cite{Khani_2021_ICCV} and model/feature caching can further improve inference efficiency by selectively invoking larger models or reusing cached embeddings and prior inference results when temporal or task continuity is available \cite{10124356}.}

 {As AI models continue to scale, edge AI is further expanding to support large AI models (LAIMs), which increasingly rely on collaborative execution and dynamic workload dispatching \cite{lyu2026quantizationawarecollaborativeinferencelarge}. In particular, several emerging architectures are becoming increasingly relevant to LAIM edge deployment. Sparse mixture-of-experts (MoE) architectures improve scalability by activating only a small subset of experts for each query, thereby reducing the effective computation and memory burden \cite{10906629}. Speculative decoding further accelerates autoregressive generation by coupling a lightweight draft model with a stronger verification model \cite{li2026flexspecfrozendraftsmeet}. In addition, prefill-decode (PD) disaggregation decouples the compute-intensive prefill stage from the memory-bound decoding stage, enabling phase-specific resource allocation \cite{jiang2025hexgen}. More broadly, retrieval-augmented generation (RAG) and agent-empowered pipelines emerge as promising extensions that couple perception, external memory, and task planning, thereby supporting hierarchical reasoning and decision making for more complex embodied and multi-modal applications \cite{feng20256gnativeaiedgenetworks}.}

 {Despite the increasing diversity of edge learning and inference paradigms, communication and computation remain the common bottlenecks that fundamentally constrain edge AI performance. This observation motivates integrated communication-computation (ICC) design as a key enabling principle for edge AI. Among the various ICC techniques, two classes are particularly relevant to edge perception in this survey: over-the-air computation (AirComp), which enables efficient aggregation of sensing data and model updates, and semantic/task-oriented communication, which supports task-aware representation exchange for collaborative inference.}

\subsubsection{AirComp}

\begin{figure*}[h]
    \centering 
	\subfigure[]{\includegraphics[width=.32\textwidth]{ 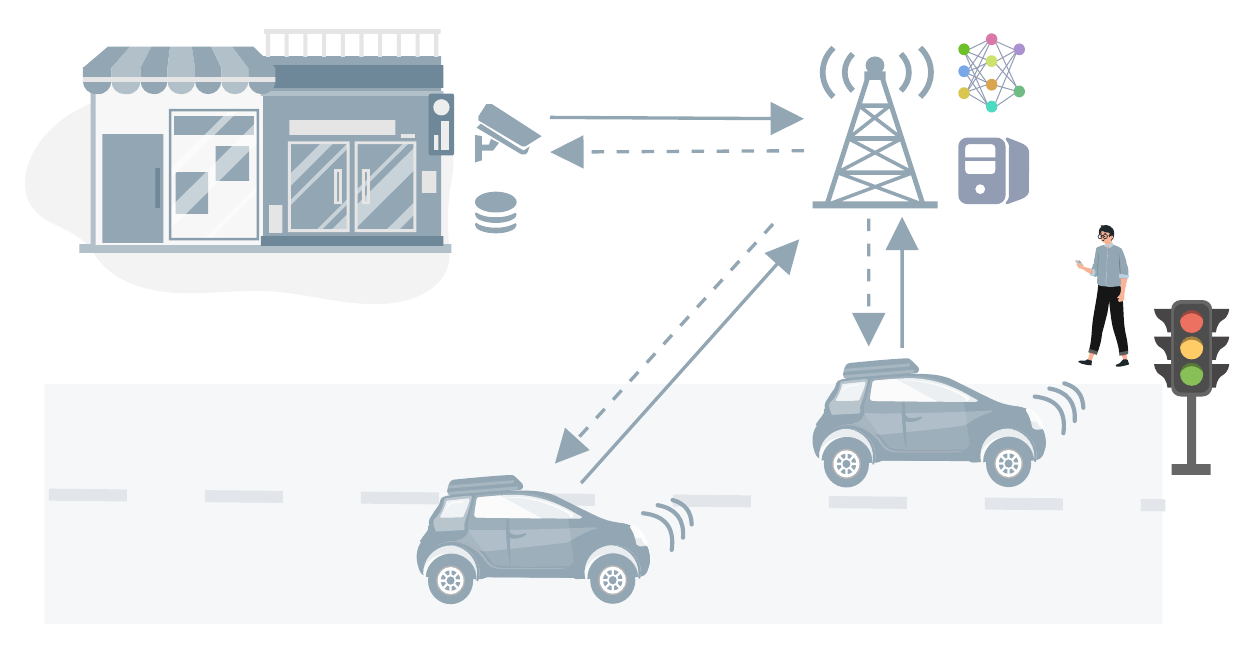}}
    \subfigure[]{\includegraphics[width=.32\textwidth]{ 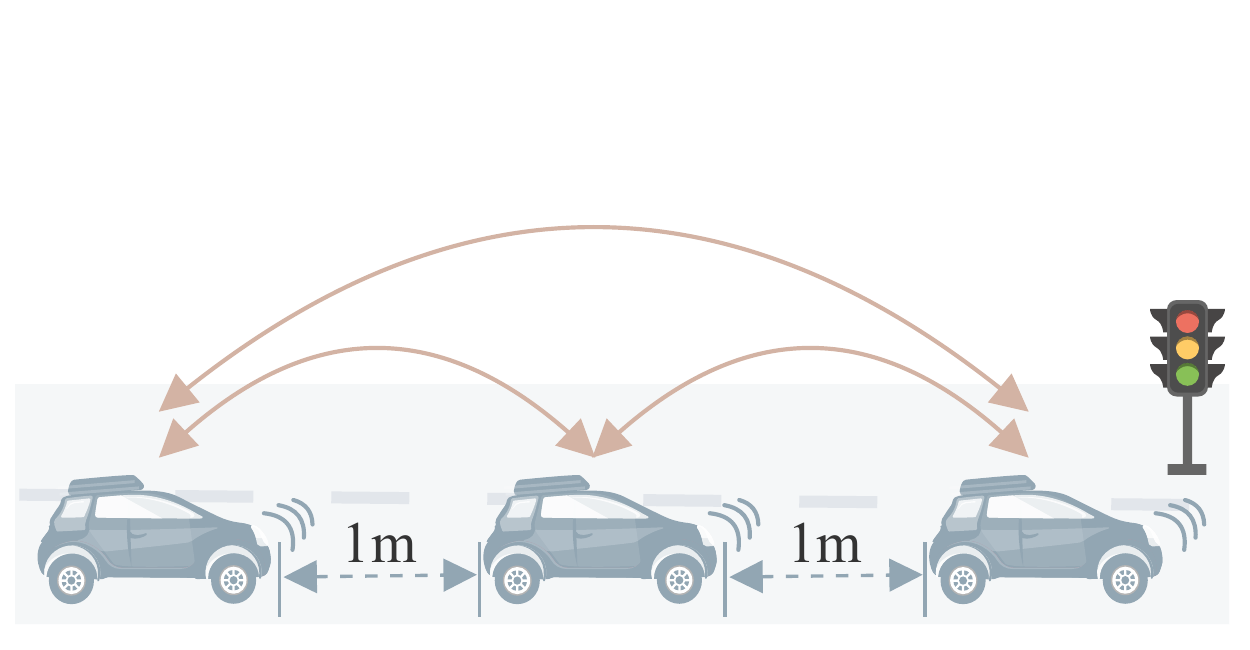}}
    \subfigure[]{\includegraphics[width=.32\textwidth]{ 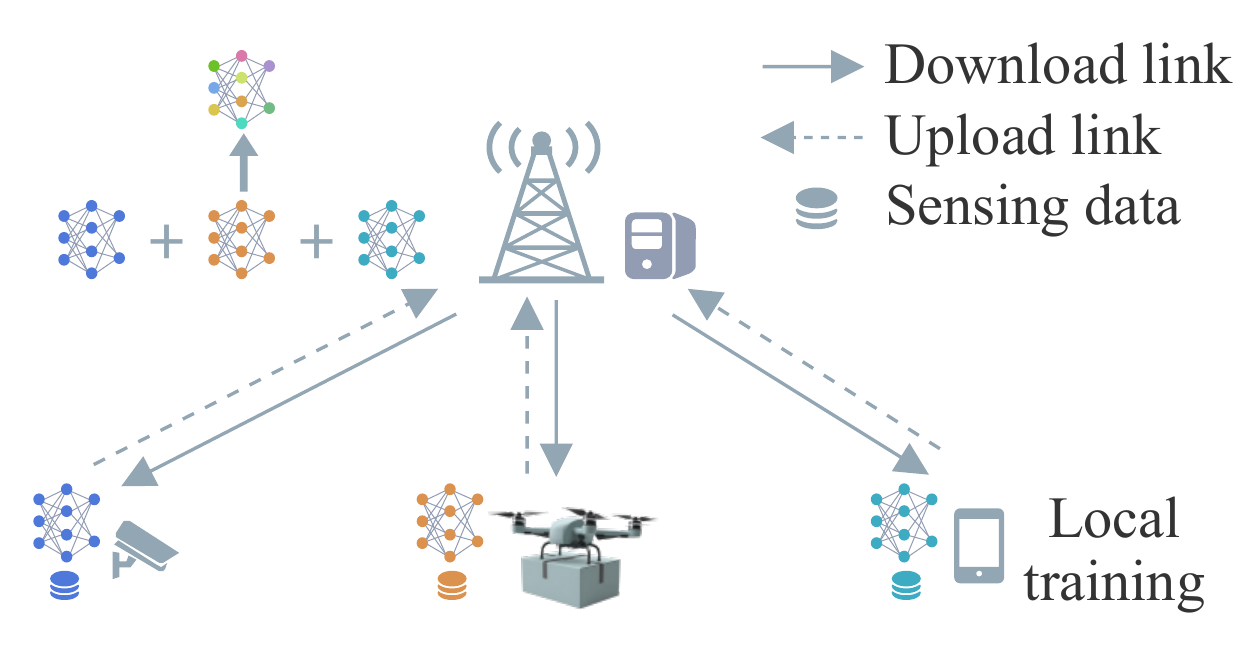}}
    \caption{ {Representative AirComp-enabled paradigms in edge perception systems, including (a) distributed sensing with over-the-air aggregation of measurements, (b) distributed consensus through analog signal superposition, and (c) collaborative edge learning with low-latency aggregation of local model updates.}}
    \label{Fig:AirComp}
\end{figure*}

AirComp is a promising ICC paradigm for facilitating function-level computation in edge AI \cite{Cao-ComM}. Unlike conventional multiple access schemes, which decode multi-user information separately in a communication-computation separation paradigm, AirComp leverages the signal superposition property of the uplink multiple access channel to perform function execution over the air \cite{GX18IoT,Cao19TWC}. 
It holds significant application potential in scenarios like distributed sensing, learning, and consensus, as shown in Fig. \ref{Fig:AirComp}.

The concept of AirComp was first introduced in \cite{Nazer07TIT}, showing that distributed devices can exploit wireless interference for reliable function computation via structured codes. Later, \cite{Gastpar08TIT} proved that uncoded analog transmission achieves minimum distortion for i.i.d. Gaussian sources, whereas coded transmission is required for correlated cases \cite{Tavildar08TIT, Vishwanath12TIT}.
From a signal processing perspective, analog AirComp has been extensively studied. In \cite{JXiaoTSP08_AirC}, an optimal linear decentralized estimation scheme was proposed for static channels, while \cite{CWangTSP11_AirC} analyzed the distortion outage performance under power and bandwidth constraints. 
The role of CSI has also been widely investigated. Specifically, \cite{Goldenbaum14WCL} shows that single-antenna systems require only channel magnitude, while robust MIMO designs under partial or unknown CSI have been studied in \cite{Dong20TSP}. The effect of channel estimation errors has been further analyzed in \cite{Chen2024TCOM}, showing that large antenna arrays can enable near error-free computation.

Unlike traditional data aggregation in sensor networks, AirComp has been applied to alleviate the communication bottleneck in federated learning (FL), leading to the paradigm of over-the-air federated edge learning (Air-FEEL) \cite{Cao-ComM}. By shifting the design objective toward learning performance, this task-driven view has motivated extensive research on learning-oriented wireless optimization. For example, \cite{KYang2020TWC} jointly optimizes user selection and receive beamforming to maximize device participation under a computation MSE constraint, while \cite{GZhu2020TWC} proposes truncated power control to exclude deeply faded devices, balancing learning performance and aggregation error.
\cite{Cao2021-FedAvg,Cao2021AirFEEL} further analyze the trade-off between update errors and learning performance, and optimized power control to enhance model accuracy.

\subsubsection{Semantic and Task-oriented Communications}

\begin{figure*}[t]
  \centering
  \includegraphics[width=0.95\textwidth]{ 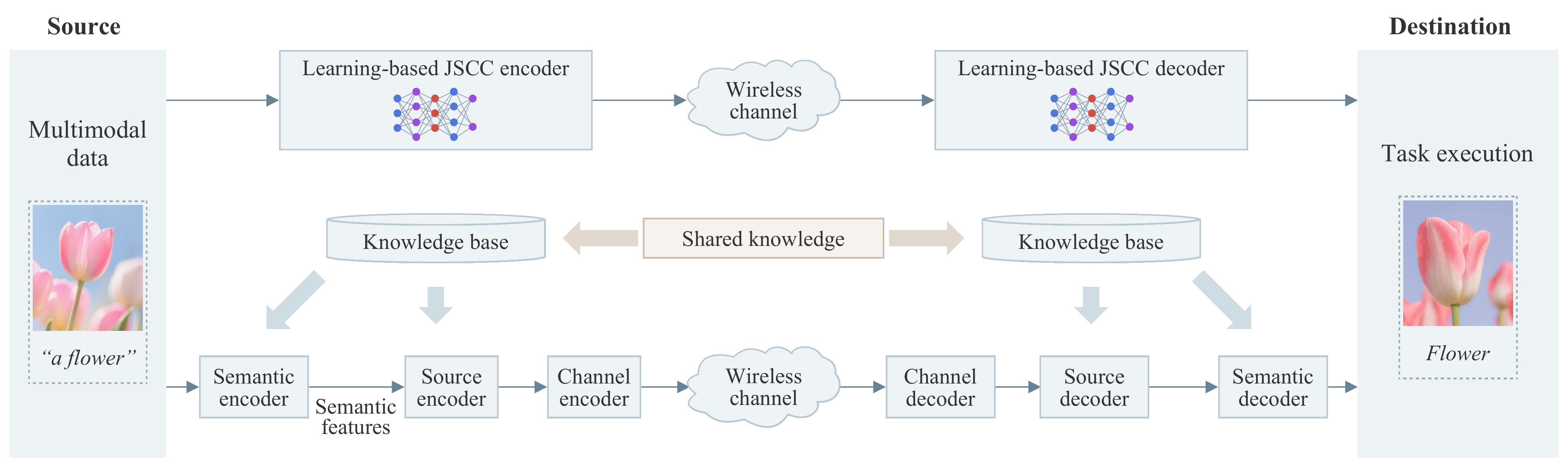}
  \caption{ {Basic workflow of semantic/task-oriented communications, where multi-modal source data are mapped to task-relevant semantic representations and transmitted over wireless channels, supported by shared knowledge bases at both the transmitter and receiver to enable reliable data recovery or task execution.} }
  \label{Fig:Procedure}
\end{figure*}

Complementary to AirComp, semantic and task-oriented communications offer another key realization of ICC for edge perception by shifting the transmission objective from symbol-level recovery to task-level information delivery \cite{Lan21JCIN}. Semantic communication focuses on preserving data meaning by minimizing semantic distortion \cite{Tilahun24IEEE}, while task-oriented communication further optimizes transmission directly for downstream tasks such as classification or detection \cite{Shi23WC}. By transmitting only task-relevant information, these approaches significantly reduce bandwidth usage and improve scalability for distributed AI systems. Their core objective is to efficiently deliver the information that matters for inference, prediction, or decision-making, without reconstructing raw data at the receiver \cite{Zhang25CST-Intellicise,9955525,11178221}.  {Two main paradigms are commonly adopted to achieve this goal, as illustrated in Fig. \ref{Fig:Procedure}.
Specifically, learning-based joint source-channel coding (JSCC) directly maps source symbols to channel symbols through end-to-end training, inspired by uncoded transmission, and jointly performs compression and error protection to achieve robust reconstruction over varying wireless channels \cite{Deniz24IEEE}. In this paradigm, channel dynamics can be incorporated directly into end-to-end training. Alternatively, a knowledge base (KB) assisted approach builds on conventional source-channel separation, where task-relevant semantics are first extracted and compressed using shared knowledge, then transmitted via standard digital source and channel coding, and finally reconstructed for downstream AI tasks \cite{Ren24WC}. In this case, channel effects are mainly handled in the channel coding and transmission design stages. While JSCC offers strong robustness by incorporating channel dynamics into end-to-end optimization, the knowledge base assisted paradigm is easier to deploy in practice due to its compatibility with existing digital communication systems. Nevertheless, both paradigms explicitly account for channel dynamics.}

The practical realization of semantic/task-oriented communication depends on advances in learning-based system design and semantic-aware resource management. Deep learning (DL) has enabled end-to-end semantic communication, where representative models such as DeepSC employ Transformer architectures to directly optimize semantic fidelity and improve robustness to channel impairments \cite{Xie21TSP}, while related deep JSCC schemes further enhance reliability by mapping source data directly to channel symbols \cite{Guo24TCOM}. These learning-based frameworks have been extended to diverse data modalities, including images, text, and speech \cite{10388062,Jiang22TCOM,Weng21JSAC,10738311}. 
 {Moreover, task-oriented communication is particularly relevant to edge perception. The encoder and decoder are trained according to downstream perception tasks to extract and transmit task-relevant features more efficiently. Specifically, for object detection, the relevant distortion is associated with missed/false detections and localization errors \cite{10902550}, e.g., mean average precision (mAP) and intersection over union (IoU); for tracking, with state inconsistency and association errors \cite{11134620}; for simultaneous localization and mapping (SLAM), with pose drift and map inconsistency \cite{charalambous2025goalorientedcommunicationmultiagentsystems}; for control-oriented applications, with control-relevant quantities, e.g., control cost and stability \cite{10330567}.}
Meanwhile, resource allocation has evolved from channel state driven designs toward semantic importance-aware strategies to improve  system efficiency \cite{Wang24TCCN, 11372929}. 

Many studies enhance semantic communication by incorporating explicit knowledge bases (KBs) to go beyond purely data-driven models. For example, \cite{Yi24TWC} integrates a shared textual KB to extract residual information and reduce symbol transmission without degrading semantic performance, while \cite{Ren24WC} proposes a generative architecture that exploits source-, task-, and channel-specific sub-KBs to reduce representation dimensionality. To improve generalization across diverse contexts, \cite{Zheng24GSC} leverages a pre-trained large speech model with residual vector quantization to construct semantic KBs at both transmitter and receiver. 


{\bf Lesson Learned:}  
Edge AI is constrained by stringent latency, bandwidth, privacy, and heterogeneous device-edge-cloud deployment, challenging conventional bit-level communication. These constraints motivate AirComp for low-latency and bandwidth-efficient collaborative learning, but practical use is hindered by synchronization overhead, imperfect channel state information, aggregation distortion, and heterogeneous device participation. Similarly, transmitting semantic or task-relevant representations instead of raw data reduces communication load and improves privacy, yet its effectiveness depends on stable task objectives and appropriate distortion measures, otherwise critical sensing information may be lost. Moreover, extracting compact and robust task-aligned representations under dynamic channels and coordinating tasks across heterogeneous tiers remain difficult. Together, these challenges call for adaptive, task- and data-aware communication designs that jointly account for sensing characteristics, channel dynamics, and downstream inference objectives to support scalable edge perception.

\subsection{Operational Framework of Edge Perception}
 {The above discussions show that edge sensing and edge AI constitute the two fundamental building blocks of edge perception. However, edge perception is not merely a straightforward combination of sensing front ends and AI models. Its practical operation depends on how these building blocks are connected, coordinated, and continuously adapted under resource and QoS constraints. From this perspective, edge perception can be understood through four coupled interfaces. The sensing interface specifies how observations are acquired and represented, including sensing modality, sampling configuration, and calibration state. The communication interface governs how sensed information is exchanged across devices, edge servers, and clouds, including data format, compression level, and transmission reliability. The AI interface describes how the exchanged information is transformed into task-relevant outputs, such as predictions and control decisions. The orchestration interface coordinates the overall system by generating control commands across sensing, communication, and AI computation modules. }

 {The performance of edge perception should be evaluated from two coupled perspectives. For edge  sensing, the key objective is sensing-task performance, such as detection accuracy, tracking quality, and scene understanding capability. For edge AI, the key objective is AI-system performance, including training or inference accuracy, delay, and energy consumption. A defining feature of edge perception is that these two perspectives are linked through closed-loop adaptation. On a small timescale, the system performs online decisions, including sensing adjustments such as frame rate, beamforming, viewpoint, and modality selection, as well as communication-computation resource allocation such as offloading, transmit rate or power control, model split point, and inference path selection. On a larger timescale, the system updates AI models for both sensing and decision making, including  control models such as reinforcement learning or diffusion-based policies, as well as world or foundation models, through training, continual learning, or policy refinement. Accordingly, edge perception can be viewed as jointly optimizing sensing, communication, and intelligence (computation and model adaptation), to maximize task utility under QoS constraints (such as latency, energy, bandwidth, reliability, and privacy), while continuously updating future decisions according to current outputs and feedback signals.}

\section{Edge AI Empowered Sensing}

This section surveys how edge AI empowers different sensing modalities to deliver perception services under practical edge constraints. While various existing works adopt similar learning backbones (e.g., convolutional neural networks (CNNs) and recurrent neural networks (RNNs)), the design of edge AI for sensing is fundamentally shaped by modality-specific signal representations, task objectives, and deployment constraints. Accordingly, we discuss how sensing signals are represented, how AI models are tailored to these representations, and how they are deployed across edge tiers. We first review edge AI for in-band sensing, followed by out-of-band sensing modalities, and their cross-modality fusion.

\subsection{Edge AI for In‑band Sensing}

In-band sensing systems must operate under strict radio resource constraints, which makes edge AI a key enabler. 

\subsubsection{Radar-based Sensing} 
{\bf Signal Representations \& AI Models:} In FMCW/mmWave radar sensing, edge AI model design is strongly influenced by the structure of radar measurements, which are often represented as range-Doppler, range-angle tensors, or sparse point clouds.
AI models for radar sensing have evolved from classical approaches (e.g., support vector machines (SVMs)) operating on handcrafted features \cite{10554983} to DL-based methods, which enable the direct learning of task-relevant representations from raw or pre-processed radar data. 
For example, CNNs are widely adopted by treating radar tensors as image-like inputs, allowing convolutional filters to capture spatial patterns such as micro-motions and multipath signatures \cite{Meng_Fu_Yan_Liang_Zhou_Zhu_Ma_Liu_Yang_2020}.
For temporal applications such as gesture recognition and object tracking,  RNNs capture sequential dependencies within radar signals, enabling reliable inference in dynamic environments \cite{9305931}. 
 {Generative models further address data scarcity by synthesizing realistic radar samples \cite{9796940}.}
Beyond these mainstream DL architectures, several specialized network designs have been explored for radar perception. For example, PointNet architectures directly process radar point clouds, exploiting non-image-based representations \cite{Qi_2017_CVPR}. 
Most recently, LAIMs \cite{11301737} have been introduced to radar perception, exemplified by Radar-LLM \cite{lai2025radarllmempoweringlargelanguage},  which demonstrates the potential of foundation models for  motion understanding and semantic reasoning.

 {{\bf Edge Deployment:} Despite significant accuracy gains enabled by deep learning, radar-based edge perception is fundamentally constrained by model complexity and hardware execution capability.
For example, \cite{10188595} deploys ensemble-based tree classifiers for carry-object detection on resource-constrained Raspberry Pi and NVIDIA Jetson platforms, while \cite{10570182} achieves posture classification via a quantized tiny-ML model operating on three-dimensional (3D) radar point clouds with minimal overhead. Furthermore, hardware-centric optimization, such as Google Soli \cite{10.1145/2897824.2925953}, further illustrates how algorithm-hardware integration enables high-throughput radar perception at the edge.
Collectively, these works highlight a recurring design mechanism, where radar representations guide model architecture, and edge deployment constraints drive lightweight and resource-efficient implementations.}

\subsubsection{Cellular-based Sensing}

{\bf Signal Representations \& AI Models:}  For cellular sensing, edge AI model design is closely tied to waveform- and channel-based signal representations, such as pilot-based channel estimates and I/Q measurements. These representations motivate learning-enabled receiver architectures. Representative examples include deep-learning-powered ISAC receivers for sensing-integrated orthogonal frequency-division multiplexing (OFDM) waveforms, which achieve robust joint data recovery and range-velocity estimation under Doppler effects, phase noise, and multi-target conditions \cite{9967989}. Model-driven receiver designs, such as ISAC-NET, further embed structured iterative signal processing into data-driven learning frameworks, thereby enhancing both demodulation reliability and sensing accuracy compared with conventional receiver architectures \cite{10474422}.

{\bf Edge Deployment:} Beyond receiver-side inference, a key design characteristic of cellular sensing lies in its tight coupling with radio access network (RAN) operation. Recent AI-RAN convergence efforts advocate a unified platform in which RAN functions and AI workloads coexist over shared edge infrastructure \cite{11159494}. By exposing radio measurements and control interfaces via the near-real-time RAN intelligent controller (near-RT RIC), AI-RAN enables sensing-oriented intelligence to be embedded natively into the RAN control plane. This allows sensing tasks to be executed close to radio control loops, enabling fast reaction and closed-loop adaptation. A representative example is SenseORAN, where a machine-learning-based radar detection extensible application (xApp) processes I/Q-derived spectrograms at the near-RT RIC to identify  radar signals and reconfigure the gNB for interference mitigation \cite{10353027}.

 {Building upon such deployment architectures, edge AI further strengthens edge-perception coupling by linking sensing outcomes with communication and computation adaptation. Continual learning mechanisms enable edge nodes to update sensing models across heterogeneous CSI domains while mitigating catastrophic forgetting under strict memory constraints \cite{hu2025cross}. In parallel, model-based online learning frameworks exploit prior structural knowledge of communication and sensing models to actively optimize waveforms and resource allocation with higher sample efficiency than purely model-free approaches \cite{10494366}. Moreover, deep reinforcement learning (DRL) has been employed to adapt bandwidth and power allocation in multi-user ISAC systems under time-varying demands and resource conditions \cite{10622875}. These mechanisms show that in cellular sensing, AI is not only deployed at the edge for perception, but also continuously feeds back into radio configuration and resource control, thereby forming a closed loop between signal acquisition, inference, and network adaptation. Such closed-loop designs naturally support task-level cellular perception, including contactless multi-user presence and activity detection via 5G CSI \cite{ashleibta20215g}, as well as high-precision localization based on uplink channel measurements in 5G new radio (NR) systems \cite{10741343}.}

\subsubsection{WLAN-based Sensing}
{\bf Signal Representations \& AI Models:}  In WLAN-based sensing, edge AI design is shaped by the high-dimensional and time-varying nature of CSI, as well as the stringent latency and energy constraints of embedded platforms.
Practical deployment relies on efficient and lightweight signal processing pipelines that extract amplitude, phase, and Doppler features from CSI within microsecond-to-millisecond latency budgets. These capabilities are increasingly supported by low-cost and resource-constrained edge devices, such as the ESP32 microcontroller \cite{9217780}, which can collect CSI and run compact AI models locally without relying on high-power computing platforms.  {Moreover, dense and low-cost WiFi sensing networks have been prototyped in real-world settings \cite{9525003}, where a 16-node mesh deployment can achieve 25.3\% area coverage, compared with only 4.4\% for conventional sensor deployment, thereby highlighting the scalability advantage for precision agriculture.}

 {{\bf Edge Deployment:}  On top of these representations, AI model design for WiFi sensing must jointly account for recognition accuracy, deployment efficiency, and environment-driven adaptation at the network edge. Two major directions have been developed to enable fast on-edge inference, namely lightweight architectures and model compression. On the one hand, lightweight networks, such as SqueezeNet \cite{iandola2016squeezenetalexnetlevelaccuracy50x} and MobileNet \cite{howard2017mobilenetsefficientconvolutionalneural}, significantly reduce parameter counts and computational burden, thereby enabling real-time inference under stringent memory and compute budgets. On the other hand, compression techniques, including structured pruning and post-training quantization \cite{9622251}, further shrink model size with limited accuracy loss. Beyond efficient inference, a key challenge of WLAN sensing lies in the tight coupling between CSI dynamics and model robustness under changing environments. In particular, environmental variation directly alters CSI distributions, which requires sensing models to be continuously adapted rather than statically deployed. To this end, \cite{10.1145/3408308.3427983} and \cite{9659826} propose on-device learning and federated edge learning pipelines that update WiFi sensing models using streaming CSI, thereby improving resilience against environmental changes while preserving user privacy.}

These advances in efficient pre-processing and lightweight model design have enabled a broad range of on-edge WiFi sensing applications. A major application is localization, where target positions are inferred in device-based or device-free manners using CSI fingerprints \cite{9161004}.  Human activity recognition constitutes another core task, in which stationary and mobile activities are identified from temporal CSI dynamics for fall detection \cite{8391737} and gesture recognition \cite{10417100}.  Beyond individual-level perception, crowd counting and occupancy detection leverage CSI fluctuations to estimate crowd sizes and monitor occupancy in public or secure spaces, offering privacy-preserving analytics compared with vision-based approaches \cite{10637758}.   Finally, emerging applications further demonstrate the multi-functionality of edge-enabled WiFi sensing, including contactless health monitoring \cite{10379134} and agricultural sensing tasks such as fruit ripeness estimation \cite{10552143}.

\subsubsection{Other In-band Sensing Modalities}
Other in-band modalities, such as radio frequency identification (RFID), Bluetooth, and long-range radio (LoRa), also benefit from edge AI. 
These modalities rely on lightweight measurements such as RSSI/phase, packet-level features, or symbol-level statistics. For example, in RFID sensing, device-free gesture recognition can be achieved by processing received signal strength and phase with lightweight classifiers (e.g., random forest and SVM) deployed on embedded platforms \cite{9804803}. 
For Bluetooth, BlueSeer \cite{10.1145/3489517.3530519} shows that environmental context can be inferred solely from received Bluetooth packets with embedded NNs on mobile devices. Moreover, LoRa sensing has been explored for human gait recognition, where LoGait extracts gait profiles from communication symbols via dedicated pre-processing and dynamic time warping-based learning \cite{10194558}.

\subsection{Edge AI for Out‑of‑band Sensing}
Compared with in-band wireless sensing, out-of-band sensing modalities typically offer richer semantic and geometric information, but also pose significant challenges in terms of data volume, modality heterogeneity, and deployment cost. Edge AI plays a pivotal role in addressing these challenges, as summarized in Table \ref{tab:edge_ai_empowered_sensing}.

\subsubsection{Vision-based Sensing}
{\bf Signal Representations \& AI Models:} 
From a representation perspective, edge vision systems must cope with the high dimensionality and data rates of visual streams, which pose significant challenges to bandwidth- and latency-constrained deployments. As a result, efficient feature extraction and redundancy reduction are essential to minimize transmission overhead prior to inference. Typical operations include frame sampling and cropping, which skip non-informative frames and isolate regions of interest (RoIs) with high semantic value for downstream analytics \cite{10.1145/3495243.3517016}. More recently, neural feature compression methods \cite{10540267} have extended this paradigm by deploying lightweight encoders at edge devices to extract task-aware feature representations, thereby further reducing bandwidth consumption without significantly compromising accuracy.

 {{\bf Edge Deployment:}  Building upon compact visual representations, edge AI models are designed to support real-time and accurate visual inference under dynamic workloads. Profiling is a crucial step to quantify the trade-off between computational demand and inference accuracy, thereby identifying efficient configurations under time-varying workloads \cite{8567661}. Beyond profiling, hierarchical inference, such as early-exit networks \cite{9769868} and model cascades \cite{Khani_2021_ICCV}, has become a dominant paradigm for low-latency edge vision perception, where computation depth is adaptively adjusted according to frame complexity and task urgency. Complementarily, semantic communication frameworks \cite{10388062} transmit abstract feature representations rather than raw pixels, enabling task-oriented inference with substantially reduced transmission overhead. Together, these mechanisms show that visual sensing and AI inference are increasingly coupled for edge perception.}

 {From a deployment and system optimization perspective, scalable architectures have been explored to coordinate cameras, edge servers, and cloud resources. For instance, \cite{10.1145/2789168.2790123} designs a server-based wireless video surveillance system that processes analytics tasks on edge servers connected to cameras, where only semantically relevant frame segments are uploaded to a central controller. To fully exploit the cooperative potential of heterogeneous computing nodes, \cite{8695132} proposes a distributed object detection architecture, where edge devices, servers, and clouds collaborate for coordinated visual inference. In parallel with architecture design, system-level resource management is essential for real-world deployment. Adaptive configuration tuning mechanisms dynamically adjust camera parameters to approach Pareto-optimal trade-offs between inference accuracy and resource efficiency \cite{10884674}, while task offloading determines analytics workload scheduling under varying latency and bandwidth conditions \cite{10025689}. Beyond compute scheduling, model caching and storage strategies \cite{10138921} improve scalability by coordinating cache updates, thereby reducing model migration costs and service interruptions. Overall, vision-based edge perception is evolving from isolated edge processing toward cooperative, hierarchical, and resource-aware systems by orchestrating representations, inference models, and system optimization under edge constraints.}

\subsubsection{LiDAR-based Sensing}
{\bf Signal Representations \& AI Models:}  LiDAR point clouds are inherently unstructured and voluminous, posing significant challenges for real-time processing on resource-constrained edge platforms. From a representation perspective, LiDAR sensing relies on sparse and irregular 3D point clouds, which are typically processed by detection or semantic segmentation models operating on point-based or voxelized representations. These characteristics make LiDAR perception both computation- and memory-intensive, motivating careful model and system co-design at the edge.

 {{\bf Edge Deployment:}  LiDAR-based edge perception is fundamentally shaped by the tight coupling among point-cloud representation, communication overhead, and edge AI execution. On the model side, recent studies have evaluated deep 3D perception networks under embedded constraints. For example, \cite{haidar2024readyrealtimelidarsemantic} benchmarks LiDAR semantic segmentation models on NVIDIA Jetson platforms and identifies the key computation and memory bottlenecks that hinder real-time execution on embedded hardware. To alleviate these limitations, edge-assisted inference has been explored, where point-cloud processing is selectively offloaded to edge servers to improve latency and energy efficiency \cite{10.1145/3539491.3539591}. This design highlights a fundamental coupling in LiDAR-based edge perception: the sensing representation, communication load, and inference placement must be jointly coordinated, rather than optimized in isolation. Furthermore, joint optimization of data representation and system deployment plays a crucial role in enabling scalable LiDAR-based edge perception. Point-cloud compression has been investigated to reduce storage and transmission overhead. For instance, \cite{8949733} introduces a teacher-student learning paradigm for LiDAR map compression that preserves localization accuracy. In addition, collaborative edge-cloud architectures further strengthen such coupling by coordinating perception across devices, edges, and infrastructure.  {For example,  \cite{10588634} develops an edge-based micro-pulse LiDAR system for efficient vehicle recognition.  \cite{10815971} demonstrates a real-time roadside LiDAR AIoT prototype that achieves high accuracy of 96.03\%  with an average latency of 11.78 ms, and the resulting edge-cloud-terminal pipeline reduces total latency from 871.02 ms to 68.39 ms by compressing the point-cloud stream from 37 MB/s to 2.4 MB/s.} Also, EdgeCooper \cite{10274112} proposes an edge-assisted multi-vehicle cooperative perception framework that extends perception range.}

\begin{figure}[t]
  \centering
  \includegraphics[width=0.5\textwidth]{ 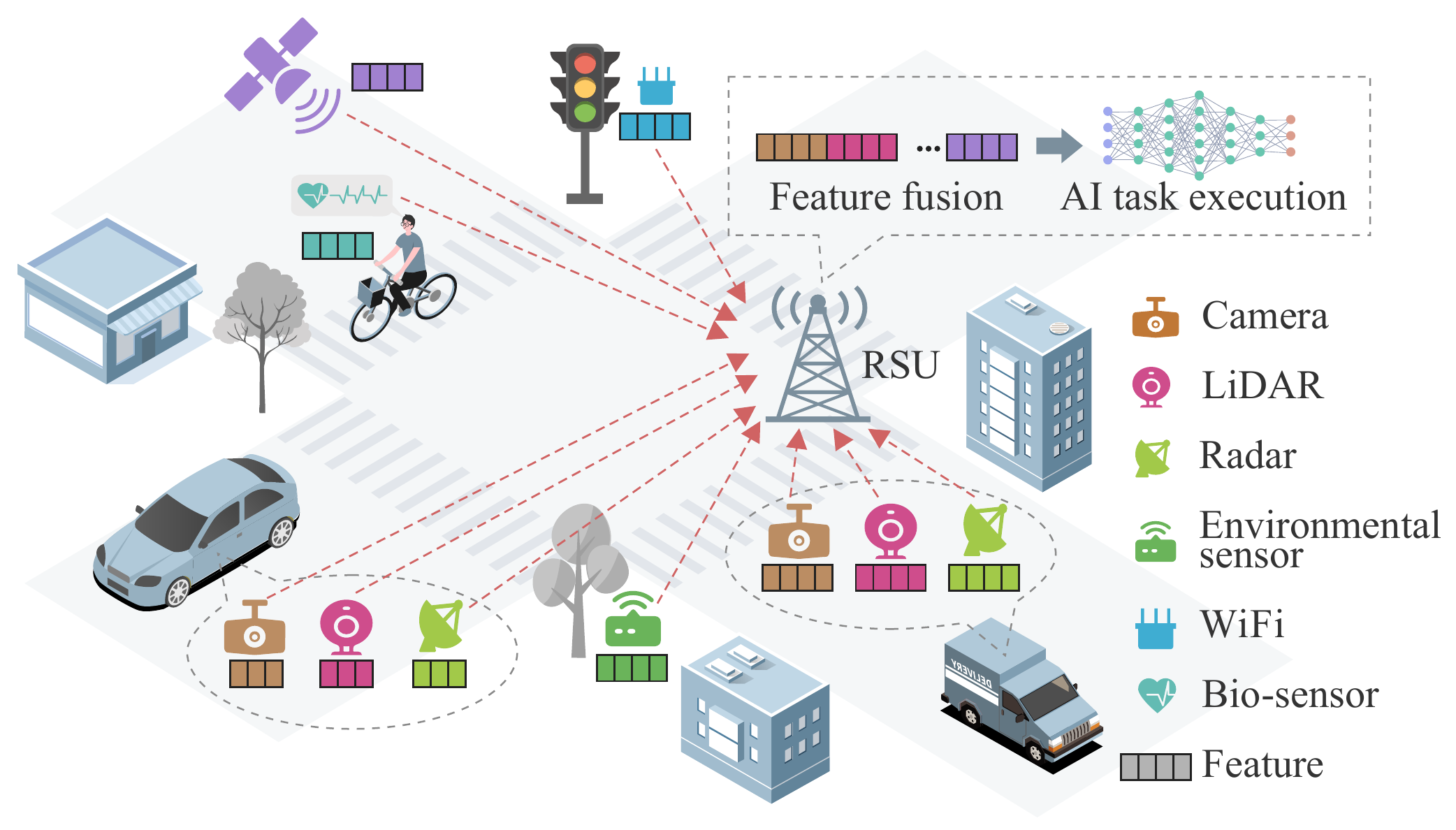}
  \caption{ {Illustration of multi-modal feature fusion in edge perception, where heterogeneous features extracted from in-band and out-of-band sensing modalities are aggregated at the network edge to enhance downstream AI task execution.}}
  \label{Fig:fusion}
\end{figure}

\subsubsection{Environmental Sensing}

{\bf Signal Representations \& AI Models:}  Environmental sensing plays a fundamental role in monitoring air quality, water resources, infrastructure health, and large-scale ecological systems. Environmental data are typically sparse, noisy, and long-term, and are collected by resource-constrained IoT sensors under limited power, memory, and connectivity. These characteristics make edge AI particularly well suited for environmental perception, as it enables event-triggered reporting, low-latency responses, and privacy protection  \cite{11125738}. In the following, we discuss how edge AI addresses key challenges across representative environmental sensing modalities.

 {{\bf Edge Deployment:} By jointly leveraging low-power sensing technologies, lightweight AI models, and hierarchical coordination, edge AI elevates environmental perception from passive data collection to fast, proactive, and sustainable environmental intelligence.}

 {First, air and water quality monitoring often relies on dense deployments of low-cost sensors, which may suffer from drift, calibration errors, and communication bottlenecks. With edge AI, local anomaly detection and adaptive calibration can be performed near sensors, reducing backhaul traffic and shortening response time. For instance, \cite{9354565,loukatos2022enriching} develop edge AI-enabled air-quality and water-monitoring systems in which distributed sensors execute real-time inference and issue local notifications without continuous cloud dependence.
Second, acoustic perception is increasingly used for environmental awareness in urban and natural ecosystems, such as fire alarms, wildlife monitoring, and smart-city noise management. In this case, edge AI supports context-aware edge perception through on-device feature extraction and noise classification, enabling rapid detection of abnormal acoustic events while protecting privacy by avoiding raw-audio transmission \cite{szymoniak2025overview}. 
Third, mechanical sensors (e.g., vibration and pressure) are essential for structural health monitoring and fault detection. For example, \cite{11096250} implements a field programmable gate array (FPGA)-based edge AI framework where quantized neural networks perform vibration classification directly on IoT nodes with reduced inference time and power consumption, highlighting the benefit of hardware-algorithm co-optimization for continuous monitoring.
Fourth, advances in nano- and micro-scale sensors are pushing environmental perception toward finer-grained and self-sustained intelligence. The intelligent nano-microsystems in \cite{guo2025advances} integrate energy harvesting, photonic transduction, and edge-AI computation into compact platforms capable of localized chemical and molecular inference without external power. These innovations enable self-powered and ultra-precise environmental perception even in harsh or inaccessible conditions.
Overall, the integration of edge AI into environmental sensing has enabled a broad range of applications, including smart industrial monitoring \cite{8658105} and precision agriculture \cite{el2024leveraging}. }

\subsubsection{Biosensing}
{\bf Signal Representations \& AI Models:}  Recent progress in wearable bioelectronics and edge AI has shifted intelligent biosensing from cloud-centric pipelines toward distributed edge architectures comprising biosensors, mobile devices, and nearby edge servers \cite{bios15070410}. Biosignals, such as electrocardiograms, respiration, motion, and biomedical images, are typically continuous, personalized, and privacy-sensitive, which, together with stringent latency requirements, motivates near-sensor edge intelligent processing.

\begin{table*}[h]
\centering
\caption{Comparison of Edge-AI-Empowered Sensing Paradigms Across Heterogeneous Modalities}
\label{tab:edge_ai_empowered_sensing}
\setlength{\tabcolsep}{2pt}
\renewcommand{\arraystretch}{1.4}
\fontsize{7.5}{7.5}\selectfont
{
\begin{tabularx}{\textwidth}{
>{\raggedright\arraybackslash}p{1.2cm}
>{\raggedright\arraybackslash}p{3.1cm}
>{\raggedright\arraybackslash}p{3.5cm}
>{\raggedright\arraybackslash}p{3.5cm}
>{\raggedright\arraybackslash}p{3cm}
>{\raggedright\arraybackslash}X}
\toprule
\multicolumn{1}{c}{\textbf{Modality}} 
& \multicolumn{1}{c}{\textbf{Signal representation}} 
& \multicolumn{1}{c}{\textbf{Edge AI model}} 
& \multicolumn{1}{c}{\textbf{Edge deployment pattern}} 
& \multicolumn{1}{c}{\textbf{Design goal}} 
& \multicolumn{1}{c}{\textbf{Main bottleneck}} \\
\midrule

\textbf{Radar-based sensing}
& Range-Doppler / range-angle tensors and sparse point clouds \cite{10554983,lai2025radarllmempoweringlargelanguage}
& SVMs on handcrafted features \cite{10554983}, CNNs \cite{Meng_Fu_Yan_Liang_Zhou_Zhu_Ma_Liu_Yang_2020}, RNNs \cite{9305931}, generative models \cite{9796940}, PointNet-style point-cloud processing \cite{Qi_2017_CVPR}, and LAIM/LLM-based reasoning \cite{11301737,lai2025radarllmempoweringlargelanguage}
& Raspberry Pi / NVIDIA Jetson deployment with ensemble trees \cite{10188595}, quantized tiny-ML on 3D point clouds \cite{10570182}, and hardware-centric co-design such as Soli \cite{10.1145/2897824.2925953}
& Gesture, fall, posture, carry-object, and motion understanding \cite{9305931,10188595,10570182,lai2025radarllmempoweringlargelanguage}
& Embedded compute/memory constraints and the need for lightweight, hardware-aware implementations \cite{10188595,10570182,10.1145/2897824.2925953} \\

\textbf{Cellular-based sensing}
& Pilot-based channel estimates \cite{9967989}, I/Q measurements / spectrograms \cite{10353027}, and uplink channel measurements in 5G NR \cite{10741343}
& DL-based ISAC receivers \cite{9967989}, model-driven learning such as ISAC-NET \cite{10474422}, continual learning \cite{hu2025cross}, model-based online learning \cite{10494366}, DRL-based resource adaptation \cite{10622875}, and attention-aided localization \cite{10741343}
& AI-RAN / O-RAN integration \cite{11159494}, near-RT RIC / xApp deployment via SenseORAN \cite{10353027}, and closed-loop sensing-resource adaptation \cite{hu2025cross,10494366,10622875}
& Joint data recovery and range-velocity estimation \cite{9967989}, radar/interference detection \cite{10353027}, presence/activity detection \cite{ashleibta20215g}, and localization \cite{10741343}
& Tight coupling with RAN control loops, heterogeneous CSI domains, and time-varying radio/resource conditions \cite{hu2025cross,10494366,10622875,11159494} \\

\textbf{WLAN-based sensing}
& CSI amplitude / phase / Doppler features \cite{9217780,10.1145/3408308.3427983,9659826}
& Lightweight architectures such as SqueezeNet \cite{iandola2016squeezenetalexnetlevelaccuracy50x} and MobileNet \cite{howard2017mobilenetsefficientconvolutionalneural}, pruning / quantization \cite{9622251}, on-device learning \cite{10.1145/3408308.3427983}, and federated edge learning \cite{9659826}
& ESP32-based standalone sensing \cite{9217780}, dense low-cost WiFi sensing networks \cite{9525003}, and streaming-CSI model updates on edge devices \cite{10.1145/3408308.3427983,9659826}
& Real-time, low-power WiFi sensing and robust recognition under environmental changes \cite{9217780,10.1145/3408308.3427983,9659826}
& High-dimensional and time-varying CSI under stringent latency/energy budgets and environmental dynamics \cite{9217780,10.1145/3408308.3427983,9659826} \\

\textbf{Other in-band sensing modalities}
& RFID RSSI / phase \cite{9804803}, Bluetooth packet-level features \cite{10.1145/3489517.3530519}, and LoRa communication symbols / gait profiles \cite{10194558}
& Random forest / SVM for RFID \cite{9804803}, embedded NNs for Bluetooth sensing \cite{10.1145/3489517.3530519}, and DTW-based learning for LoRa gait recognition \cite{10194558}
& Embedded / mobile inference with lightweight feature extraction \cite{9804803,10.1145/3489517.3530519,10194558}
& Device-free gesture recognition \cite{9804803}, environment detection \cite{10.1145/3489517.3530519}, and gait recognition \cite{10194558}
& Limited measurement richness and reliance on highly lightweight preprocessing / inference pipelines \cite{9804803,10.1145/3489517.3530519,10194558} \\

\textbf{Vision-based sensing}
& Frames, RoIs, sampled / cropped visual streams \cite{10.1145/3495243.3517016}, compressed visual features \cite{10540267}, and semantic features for task-oriented transmission \cite{10388062}
& Learnable input filtering \cite{10.1145/3495243.3517016}, scalable feature compression \cite{10540267}, profiling \cite{8567661}, early-exit inference \cite{9769868}, model cascades / adaptive streaming \cite{Khani_2021_ICCV}, and semantic communication \cite{10388062}
& Hierarchical clusters \cite{8567661}, server-based surveillance \cite{10.1145/2789168.2790123}, distributed device-edge-cloud inference \cite{8695132}, camera-parameter adaptation \cite{10884674}, task offloading \cite{10025689}, and model caching \cite{10138921}
& Low-latency and bandwidth-efficient visual analytics \cite{10.1145/3495243.3517016,10540267,8567661,9769868,Khani_2021_ICCV}
& High-dimensional visual streams together with bandwidth/computation pressure and dynamic workload scheduling overhead \cite{10.1145/3495243.3517016,10540267,10884674,10025689,10138921} \\

\textbf{LiDAR-based sensing}
& Sparse / irregular 3D point clouds and point- / voxel-based representations \cite{10.1145/3539491.3539591,haidar2024readyrealtimelidarsemantic,8949733}
& 3D segmentation / detection under embedded constraints \cite{haidar2024readyrealtimelidarsemantic}, teacher-student compression \cite{8949733}, and cooperative perception \cite{10274112}
& Edge-assisted point-cloud offloading \cite{10.1145/3539491.3539591}, Jetson benchmarking \cite{haidar2024readyrealtimelidarsemantic}, roadside edge-cloud-terminal prototype \cite{10815971}, and multi-vehicle edge-assisted perception \cite{10274112}
& Real-time 3D perception, localization, and extended vehicular awareness \cite{10.1145/3539491.3539591,8949733,10815971,10274112}
& Computation- and memory-intensive point-cloud processing, plus storage / communication overhead for cooperative perception \cite{haidar2024readyrealtimelidarsemantic,8949733,10815971,10274112} \\

\textbf{Environ-mental sensing}
& Sparse, noisy, and long-term IoT measurements \cite{11125738}, air / water monitoring data \cite{9354565,loukatos2022enriching}, acoustic signals \cite{szymoniak2025overview}, vibration / pressure signals \cite{11096250}, and nano-/micro-scale sensing \cite{guo2025advances}
& Local anomaly detection and adaptive calibration \cite{9354565,loukatos2022enriching}, acoustic classification \cite{szymoniak2025overview}, quantized NNs on FPGA \cite{11096250}, and self-sustained edge-AI microsystems \cite{guo2025advances}
& Multi-modal IoT node with edge AI \cite{11125738}, distributed local notification without continuous cloud dependence \cite{9354565,loukatos2022enriching}, FPGA-based edge nodes \cite{11096250}, and self-powered localized inference \cite{guo2025advances}
& Low-power, privacy-friendly, and sustainable environmental intelligence \cite{11125738,9354565,loukatos2022enriching,11096250,guo2025advances}
& Limited power, memory, and connectivity, together with sensor drift / calibration errors and harsh deployment conditions \cite{11125738,9354565,loukatos2022enriching,guo2025advances} \\

\textbf{Biosensing}
& Electrocardiograms, respiration, motion, biomedical images, and multi-modal wearable signals \cite{bios15070410,A2024e28688,matsumura2025real,li2025towards}
& CNN-LSTM-based motion analysis \cite{A2024e28688}, multi-modal biosignal analytics \cite{matsumura2025real}, and local ML-driven wound-state analysis and actuation \cite{li2025towards}
& Wearables with Raspberry Pi / smartphone-assisted analytics \cite{A2024e28688,matsumura2025real}, distributed biosensor-mobile-edge architectures \cite{bios15070410}, and closed-loop bioelectronic actuation \cite{li2025towards}
& Real-time, personalized, and privacy-sensitive health-aware perception / decision making \cite{bios15070410,A2024e28688,matsumura2025real,li2025towards}
& Continuous and personalized biosignals under strict latency/privacy constraints, together with motion artifacts and user robustness issues \cite{bios15070410,A2024e28688} \\
\bottomrule
\end{tabularx}
}
\end{table*}

 {{\bf Edge Deployment:} Building upon these signal characteristics, edge AI models are designed to extract task-relevant physiological and behavioral patterns directly from raw or lightly processed biosignals, thereby tightly coupling continuous sensing with local inference and timely intervention.  {For example, in fall detection and prevention, \cite{A2024e28688} designs an edge AI-enabled wearable device that executes CNN-LSTM models on platforms such as Raspberry Pi for real-time motion recognition, achieving 97\% accuracy, while reducing the model computation time to 0.43656 s on Jetson Nano, compared with 2.16164 s on Raspberry Pi 4.}
For multi-modal health monitoring, the sensor patch in \cite{matsumura2025real} integrates electrocardiogram, respiration, and temperature sensing with smartphone-based edge analytics, allowing heterogeneous biosignals to be fused and interpreted on site while improving user mobility and privacy. Moreover, edge biosensing systems increasingly close the loop between local inference and adaptive decision making. For example, the a-Heal platform in personalized wound care \cite{li2025towards} combines near-sensor inference with adaptive bioelectronic actuation, where a local “ML Physician” analyzes wound images to classify healing stages and autonomously adjust drug delivery.}


\begin{table*}[h]
\centering
\caption{Operational Taxonomy of Multi-Modal Fusion in Edge Perception}
\label{tab:operational_fusion_taxonomy}
\renewcommand{\arraystretch}{1.4}
\setlength{\tabcolsep}{2pt}
{

\begin{tabularx}{\textwidth}{
>{\raggedright\arraybackslash}p{1.15cm}
>{\raggedright\arraybackslash}p{2.55cm}
>{\raggedright\arraybackslash}p{2.55cm}
>{\raggedright\arraybackslash}p{1.70cm}
>{\raggedright\arraybackslash}p{2.20cm}
>{\raggedright\arraybackslash}X}
\toprule
\multicolumn{1}{c}{\textbf{Fusion level}} 
& \multicolumn{1}{c}{\textbf{What is exchanged}} 
& \multicolumn{1}{c}{\textbf{Typical control variables}} 
& \multicolumn{1}{c}{\textbf{Main decision-maker}} 
& \multicolumn{1}{c}{\textbf{Objective metrics}} 
& \multicolumn{1}{c}{\textbf{Main practical pain points}} \\
\midrule

\textbf{Data \cite{9879243,10571852}} 
& Raw sensor measurements, e.g., images, point clouds, radar tensors, and CSI streams
& Synchronization window, sampling rate, raw-data upload ratio, and alignment policy
& Device cluster, edge server, and RSU
& Task accuracy, spatial consistency, and transmission overhead
& Strong synchronization, calibration, and coordinate-alignment dependence; highest communication and privacy burden; sensitivity to asynchronous arrivals and missing modalities \\

\textbf{Feature \cite{9709939,10843140,gunn2024lift,10637758}}
& Intermediate embeddings and BEV features
& Feature dimension, compression ratio, alignment/weighting policy, and cached history length
& Edge/cloud server
& Task accuracy, robustness, and transmission overhead
& Stale features, representation mismatch, semantic loss, and encoder-fusion co-training burden \\

\textbf{Decision \cite{9879243,9879085,10.1145/3450268.3453532}}
& Bounding boxes, track states, confidence scores, and class logits
& Confidence threshold, voting/weighting rule, and fallback modality policy
& Edge server and orchestration module
& Decision reliability, conflict robustness, and fault tolerance
& Late-stage information loss, inconsistent confidence calibration, and limited error correction capability \\
\bottomrule
\end{tabularx}

}

\end{table*}

\subsection{Multi-modal Fusion}
 {Multi-modal fusion can be broadly categorized into data fusion \cite{9879243,10571852}, feature fusion \cite{9709939,10843140,gunn2024lift,10637758}, and decision fusion \cite{9879243,9879085,10.1145/3450268.3453532}. Data fusion operates directly on raw sensing signals and offers the richest cross-modal information, but also imposes the heaviest burden on data transmission, synchronization, and privacy protection. Feature fusion aggregates intermediate representations extracted from heterogeneous modalities, and often provides a favorable tradeoff between perception performance and communication efficiency. Decision fusion combines modality-specific outputs or high-level decisions, making it generally more tolerant to delayed, degraded, or intermittently missing inputs, albeit with more limited cross-modal interaction. In practical edge perception systems, these fusion paradigms often coexist within a single pipeline. Thus, this categorization mainly indicates where fusion occurs, rather than fully capturing its main practical difficulty.}

 {From a system perspective, a more fundamental question is whether adding an extra modality is worthwhile. Multi-modal fusion can improve task performance by exploiting complementary sensing characteristics, such as richer semantics, stronger geometric consistency, and greater robustness under occlusion, adverse illumination, or challenging propagation conditions. However, each additional modality also introduces extra sensing, communication, and computation overhead, together with non-negligible burdens in synchronization, calibration, cross-modal alignment, and privacy protection. As a result, the benefit of adding modalities is not always monotonic, but depends on task difficulty, modality complementarity, and the available edge resource budget.}

 {From an operational viewpoint, the core challenge is how heterogeneous modalities are aligned, coordinated, and adaptively fused under resource and QoS constraints. In real deployments, different modalities must often be synchronized and aligned across time, space, and representation before effective fusion. These difficulties become particularly pronounced when sensing devices operate with different sampling rates, experience temporal drift, or produce measurements with heterogeneous reliability and intermittent availability. Therefore, in edge perception, multi-modal fusion should be viewed not only as an algorithmic integration problem, but also as an operational design problem as summarized in Table \ref{tab:operational_fusion_taxonomy}.}

\subsubsection{Fusion within In-band Modalities}
In-band fusion integrates wireless sensing sources to improve edge perception performance. Within single-modality systems, feature fusion is important for enhancing signal-to-noise ratio (SNR) and robustness while enriching representations without heavy uplink overhead. For example, Spatial AirFusion \cite{10843140} performs over-the-air function-level aggregation of voxel-sparse features across distributed agents, converting multiple access into waveform superposition across subcarriers to reduce aggregation error and communication cost. MSF-Net \cite{10530186} fuses cross-frequency views of WiFi CSI via an attention-based Transformer backbone, leveraging complementary coarse- and fine-grained dynamics to stabilize recognition in perturbed environments. Fusion across multiple in-band modalities further addresses heterogeneity and unequal modality contributions by jointly exploiting complementary sensing signals. For example,  MTTFNet \cite{11074286} co-optimizes RFID, WiFi, and radar through CNN-transformer hybrids with adaptive weighting and cross-modal supervision. \cite{11124438} investigates WiFi and FMCW radar fusion, showing that feature- and decision-level integration can combine WiFi's wide coverage with radar's high-resolution kinematics to improve pedestrian-flow estimation. Overall, in-band fusion is evolving from intra-modality feature enrichment (e.g., multi-frequency complementarity) toward inter-modality integration (e.g., representation alignment), exploiting time-frequency-spatial complementarities while maintaining resource efficiency and robustness.

\subsubsection{Fusion within Out-of-band Modalities}
At the algorithmic level, extensive studies have investigated multi-modal fusion among out-of-band sensors,  with architectures ranging from simple feature concatenation to attention-based fusion mechanisms (e.g., \cite{gunn2024lift}). For edge deployment, additional works emphasize lightweight design, near-sensor computation, and latency-energy trade-offs to achieve efficient perception. For example, \cite{mendez2022efficient} presents an embedded implementation of LiDAR-radar-camera perception, demonstrating a full on-edge inference pipeline for autonomous driving. Drone-mounted perception systems achieve real-time RGB-thermal fusion on NVIDIA AGX Xavier, highlighting accuracy-efficiency trade-offs under tight onboard compute budgets \cite{speth2022deep}.  On the infrastructure side, near-sensor LiDAR-camera feature extraction and communication using FPGA-integrated intelligent roadside units (RSUs) \cite{11052245} substantially reduce end-to-end perception latency while preserving fusion accuracy.

\begin{figure*}[h]
  \centering  \includegraphics[width=0.98\textwidth]{ 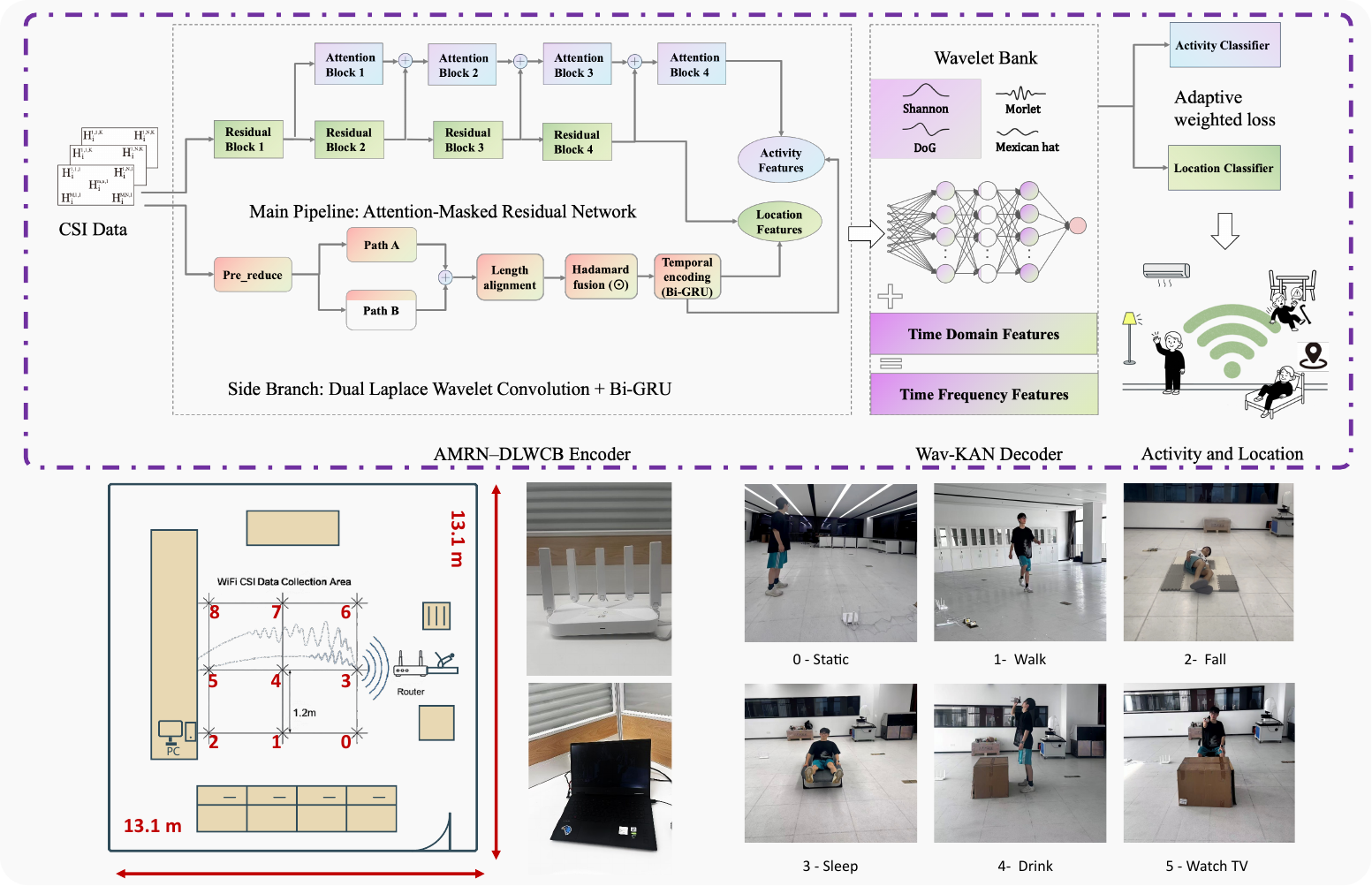}
  \caption{Case study on edge AI empowered sensing.}
  \label{Fig:AI4sensing}
\end{figure*}

\subsubsection{Fusion between In-band and Out-of-band Modalities}
Cross-domain fusion between in-band and out-of-band modalities exploits their complementary sensing characteristics: in-band sensing offers strong penetration capability, motion and velocity awareness, and robustness under adverse weather conditions, whereas out-of-band sensing provides rich semantic information and high spatial resolution. Integrating these heterogeneous sensing sources enhances edge perception reliability in challenging conditions, such as darkness, and occlusions, where single-modality systems often experience severe performance degradation,  as shown in Fig. \ref{Fig:fusion}. For example, WiFi-video fusion via a heterogeneous dual-attentional network \cite{10637758} jointly models global RF features and local visual cues, achieving improved accuracy and stability for large-scale crowd counting. MilliEye \cite{10.1145/3450268.3453532} introduces a lightweight radar-camera fusion framework with a decoupled detector, enabling real-time deployment on edge platforms and robust performance in low-light scenarios.  {For aerial vehicles, the edge-assisted Geryon system \cite{10.1145/3550298} fuses mmWave radar and camera data via multi-frame compositing and saliency-guided encoding, effectively handling low-visibility conditions and improving the mAP by 42.96\%-45.15\%,  while achieving real-time end-to-end latency of 49.5 ms and reducing bandwidth usage by 85\%-96\% compared with traditional fusion and offloading baselines.}  {Cross-domain fusion has motivated various applications, for example, in industrial settings, an edge-cloud platform combining WiFi, sub-THz imaging, radar, and infrared sensors enables passive and privacy-preserving worker monitoring around collaborative robots \cite{9146837}. Prototype results further show that feature fusion improves the average co-presence detection accuracy from 86.9\% to 97.5\%, while maintaining a stringent worker-detection latency of 90 ms.} In intelligent transportation, WaterScenes \cite{10571852} provides a multi-task radar-camera fusion benchmark for autonomous navigation, showing the robustness of joint pixel- and point-level fusion.

\subsection{Case Study and Lesson Learned}

As shown in Fig. \ref{Fig:AI4sensing}, a  multitask WiFi CSI framework is proposed in \cite{Li26TMC-DWSen}, which is built using commercial Wi-Fi equipment. In a 172m$^2$ indoor environment, a $3\times3$ grid with nine sensing locations was arranged, with a spacing of 1.2 m between adjacent positions. A ZTE AX3000 operating in the 5 GHz band, together with ZTECSITool, was used to collect CSI variations caused by human activities.

Based on this setup, the ElderAL-CSI dataset was collected. It includes data from three participants performing six daily activities at nine indoor locations, including static, walking, falling, sleeping/lying down, drinking, and watching TV. In total, 42,147 fixed-length CSI samples were obtained, and each sample was annotated with both an activity label and a location label.
This work uses a dual-path Wavelet–Attention KAN framework to jointly model the CSI data. One branch extracts informative time–subcarrier features, while the other captures multi-scale time–frequency features, enabling simultaneous activity recognition and indoor localization. Overall, the system provides a low-cost, contactless, and privacy-friendly solution for joint activity–location sensing, achieving high recognition accuracy across multiple datasets and showing strong potential for smart-home and elderly-care monitoring applications.

 {{\bf Lesson Learned:} Edge AI plays a critical role in unlocking the sensing potential across in-band and out-of-band modalities. By tailoring learning models to modality-specific signal representations, AI methods can effectively extract task-relevant structures that are difficult to capture with generic signal processing alone, thereby improving perception quality. However, these gains are tightly coupled with deployment feasibility at the network edge, which depends not only on resource-aware deployment strategies, such as lightweight model design, but also on the operational feasibility of multi-modal fusion, including synchronization, calibration, and heterogeneous sampling rates. Furthermore, multi-modal fusion emerges as a fundamental enabler for reliable edge perception, since different sensing modalities exhibit complementary strengths in semantics and robustness. Such complementarity can substantially improve task performance in adverse conditions and justify the additional fusion cost. In benign scenarios, however, the performance gain may be marginal while still incurring extra system overhead. Therefore, edge perception should adaptively select the modality set and fusion level to balance task performance against the induced overhead. Collectively, these insights highlight that high-performance edge perception requires the joint consideration of modality-aware and fusion-driven AI model design, as well as edge-aware deployment and orchestration.}

\section{Task-oriented Sensing Assisted Edge AI}
This section reviews task-oriented sensing for edge AI, where the objective is to optimize task-level AI performance, rather than classical sensing performance alone. To this end, sensing configurations, such as sensing waveforms, sampling rates, viewpoints, and sensing schedules, are jointly designed with other resources under practical edge QoS constraints.
Within this framework, ISCC and active perception  emerge as closed-loop design mechanisms, where task feedback from edge AI models is exploited to adapt sensing configurations and resource usage over time, thereby enabling efficient and selective information acquisition for downstream AI tasks, as summarized in Table \ref{tab:task_oriented_sensing_assisted_edge_ai}.

\subsection{ISCC Design}
In edge AI systems, sensing, communication, and computation constitute three tightly coupled yet functionally distinct components that jointly determine end-to-end task performance. Sensing acquires observations from the physical environment, communication transports intermediate representations among devices and edge servers, and computation executes AI inference and learning on the acquired data. In conventional network design, these components are often optimized independently, where sensing targets higher measurement quality, communication pursues greater throughput or reliability, and computation aims for faster processing speed. However, such a decoupled and task-agnostic design paradigm can give rise to  resource inefficiencies. For instance, enhancing sensing resolution may overwhelm available bandwidth, aggressive data transmission can increase energy consumption and constrain computation budgets, and excessive computation may introduce latency and degrade system responsiveness.

 {This intrinsic coupling implies that isolated optimization is fundamentally suboptimal. Accordingly, the design objective should be shifted toward task-centric AI performance metrics, such as inference accuracy, learning convergence speed, and joint energy-delay efficiency, rather than individual sensing, communication, or computation indicators. Achieving this requires task-oriented integration, in which sensing fidelity, transmission reliability, and computational load are jointly designed within a unified resource-allocation framework. The resulting ISCC paradigm transforms the conventional ``sense-then-transmit-then-compute'' pipeline into a coordinated and resource-aware loop, enabling efficient edge AI execution.}

\begin{figure}[h]

  \centering
  \includegraphics[width=0.5\textwidth]{ 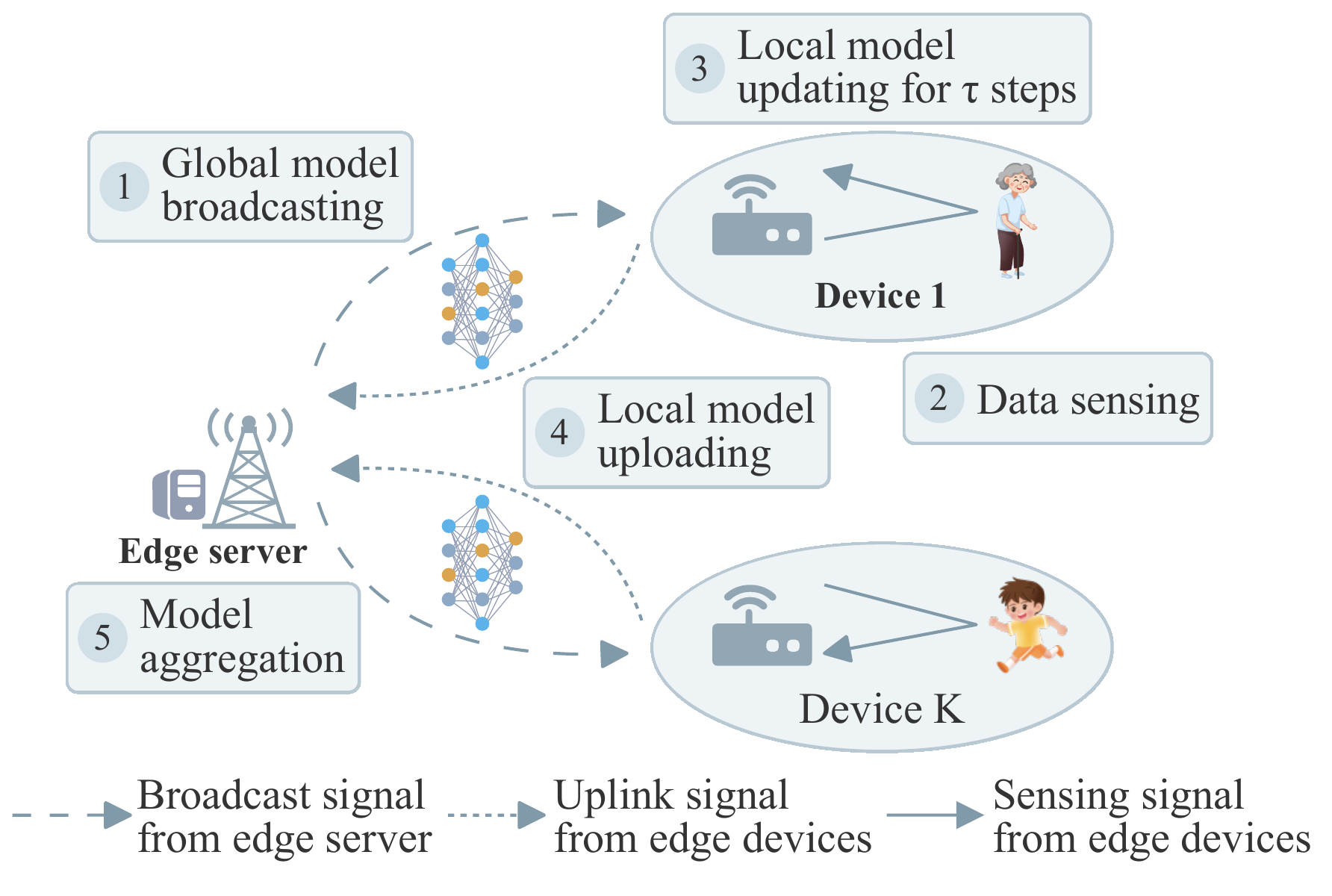}
  \caption{ {Task-oriented ISCC for federated edge learning, where sensing quality, communication latency, and learning convergence are characterized and jointly optimized to accelerate distributed model training  \cite{9970330}.}}
  \label{ISCClearning}
\end{figure}

\subsubsection{System-level Frameworks}
Recent studies have established system-level ISCC frameworks, showing how cross-domain resource coupling can be coordinated for edge AI. The work in \cite{10159270} develops a unified ISCC framework that explicitly models the competition between sensing and computation resources, and analyzes the impact of sampling rate on perception accuracy to reduce redundant sensing. Complementarily, \cite{9763441} integrates ISAC and mobile edge computing into a joint ISCC scheduling architecture, revealing multi-dimensional trade-offs among sensing, communication, and computation. By formulating device association and subchannel assignment as a matching game, the authors derive a convergent and stable coordination algorithm. Moreover, \cite{9687468} considers a 6G-oriented ISCC framework, where unified hardware and protocol co-design enables mutual reinforcement, sensing improves communication reliability, while computation feedback enhances sensing and resource allocation.

\subsubsection{ISCC for Edge Learning}
ISCC has been further exploited to improve edge learning, enabling scalable and privacy-preserving model training \cite{cao2025joint}. For example, \cite{10415206} introduces a multi-cell cooperative edge learning framework, where multiple ISAC stations jointly perform radar sensing and data offloading through sensing scheduling and dual-layer beamforming. Moreover, \cite{9970330} proposes a task-oriented ISCC resource allocation scheme for federated edge learning, as shown in Fig. \ref{ISCClearning}. By explicitly characterizing the coupling among sensing quality, communication latency, and learning convergence, the authors develop a joint optimization algorithm that significantly accelerates training under stringent QoS constraints. Collectively, these works demonstrate that ISCC offers a unified and systematic mechanism to coordinate resources across the entire learning life-cycle, thereby improving learning efficiency.

\subsubsection{ISCC for Edge Inference}
\begin{figure}[h]
  \centering
  \includegraphics[width=0.5\textwidth]{ 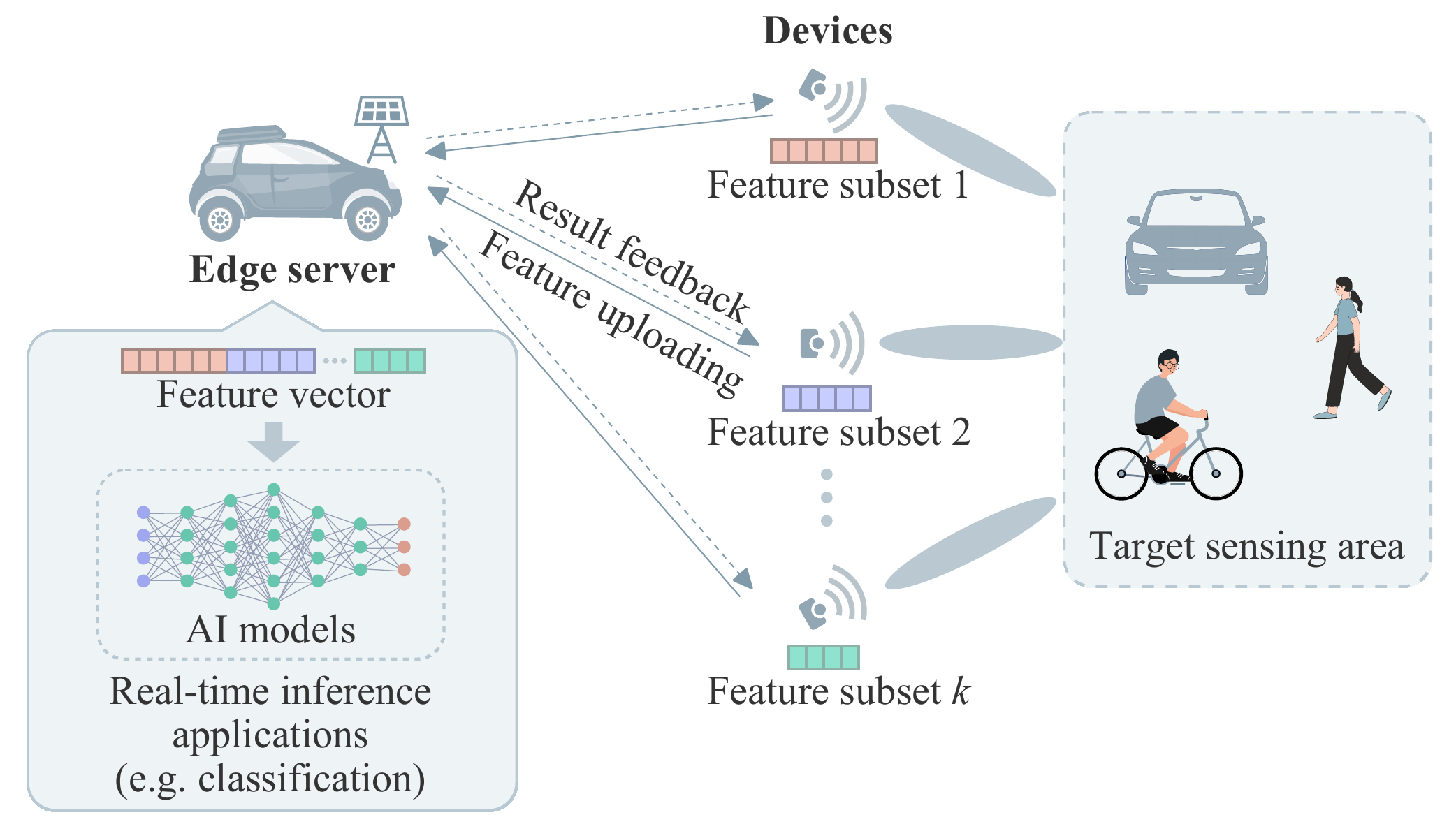}
  \caption{ {Task-oriented ISCC for multi-device edge inference, where distributed ISAC devices extract and offload sensing features to an edge server, enabling collaborative inference through the joint optimization of sensing time, transmission power, and quantization resolution  \cite{10217150}.}}
  \label{ISCCinference}
\end{figure}

The ISCC concept has also been extended to edge inference to support real-time and resource-constrained AI services. As shown in Fig. \ref{ISCCinference}, \cite{10217150} presents a task-oriented ISCC framework for multi-device inference, where distributed ISAC devices extract radar features and offload quantized representations to an edge server for model execution. The authors define a discriminant gain metric that links inference accuracy to sensing precision, quantization resolution, and communication reliability, and then derive the optimal allocation of sensing time, power, and quantization bits for collaborative inference. Building on this framework, \cite{10631278} studies mode selection among on-device, on-edge, and cooperative inference, where dual-view sensing is performed at both devices and servers. An energy-minimization algorithm subject to inference-accuracy constraints enables adaptive mode switching, and is validated on a hardware prototype platform. Together, these studies demonstrate that ISCC-aware inference facilitates  task-driven perception across heterogeneous and resource-constrained edge environments.

\subsubsection{Energy- and Delay-aware ISCC}
Beyond improving inference and learning performance of AI tasks, ISCC is also essential for energy-latency balancing in edge AI. The work in \cite{11146901} develops an energy-efficient ISCC framework that jointly optimizes sensing rate, offloading ratio, and CPU utilization across IoT devices and edge servers, and minimizes energy consumption under real-time constraints using a string-pulling algorithm and a master-slave decomposition. 
Extending to aerial networks, \cite{11318338} proposes cooperative multi-uncrewed aerial vehicles (UAVs) ISCC systems that jointly optimize sensing, trajectory, transmit power, and offloading policy for LAE services with LAIMs. These results confirm that ISCC not only enhances AI task performance, but also enables energy-efficient and low-latency edge AI operation through cross-domain optimization of resources.

\subsection{Active Perception}
\begin{figure}[h]
  \centering
  \includegraphics[width=0.5\textwidth]{ 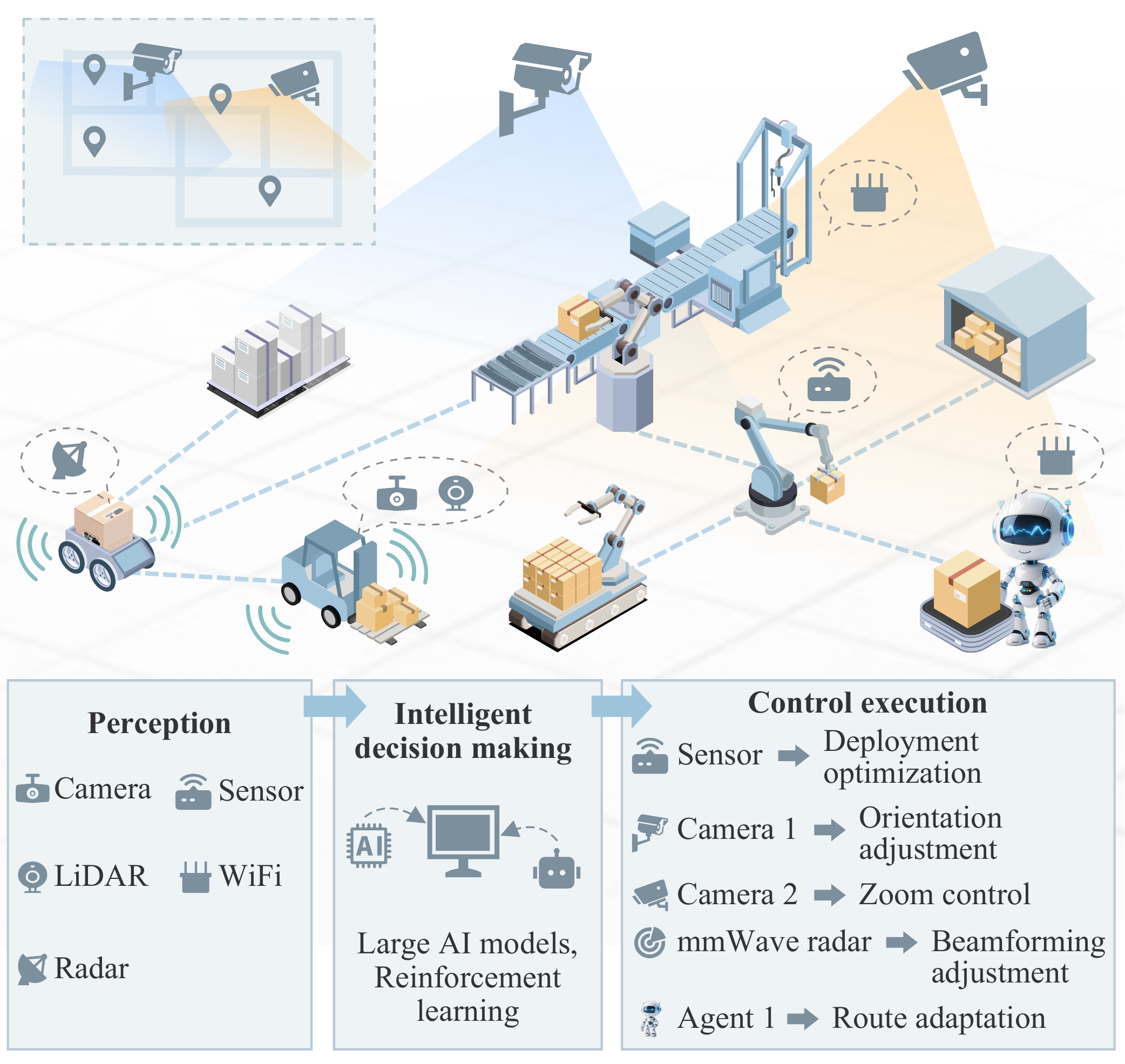}
  \caption{ {Active edge perception for downstream AI tasks, where sensing configurations (e.g., camera orientation, radar beamforming, and sensor deployment) are adaptively adjusted based on perception outcomes and task requirements.}}
  \label{Fig:activeperception}
\end{figure}

Active edge perception advances beyond blind sensing toward selective and purposeful sensing, shifting perception from passive data acquisition to task-driven decision-making. As shown in Fig. \ref{Fig:activeperception}, the key idea is to enable systems to adaptively control sensing configurations, such as camera viewpoints and spatial deployment, according to task uncertainty, environmental dynamics, and resource availability. This paradigm follows the classical perception-action coupling principle, which advocates joint optimization of perception and control to maximize AI task-relevant performance \cite{bajcsy2018revisiting}. 

\subsubsection{Task-oriented Sensing Control and Coordination}
Active perception  can be viewed along three interrelated dimensions, namely parameter adaptation, viewpoint and geometry adaptation, and multi-agent coordination, which together enable task-oriented sensing decisions in dynamic edge environments.

\textbf{Parameter adaptation:}  In edge-based video analytics and surveillance, online and adaptive adjustment of sensing parameters is essential to maintain perception robustness under dynamic illumination, scene complexity, and motion conditions. For example, CamTuner \cite{10884674} proposes a reinforcement-learning-based controller that tunes non-automated camera parameters (e.g., exposure and focus) to environmental changes, aligning sensing quality with analytics accuracy. Moving from single-camera sensing to distributed perception, CrossVision \cite{10202594} considers collaborative edge cameras observing overlapping scenes. It identifies redundant RoIs and develops a distributed on-camera scheduling strategy that balances workloads across heterogeneous nodes, avoiding repeated processing of shared views. Similarly, Argus \cite{10682605} proposes a cross-camera video analytics framework for multi-target tracking by exploiting spatio-temporal association across overlapping fields of view, dynamically prioritizing object inspections, and distributing processing among smart cameras. 

\begin{figure*}[h]
  \centering  \includegraphics[width=0.98\textwidth]{ 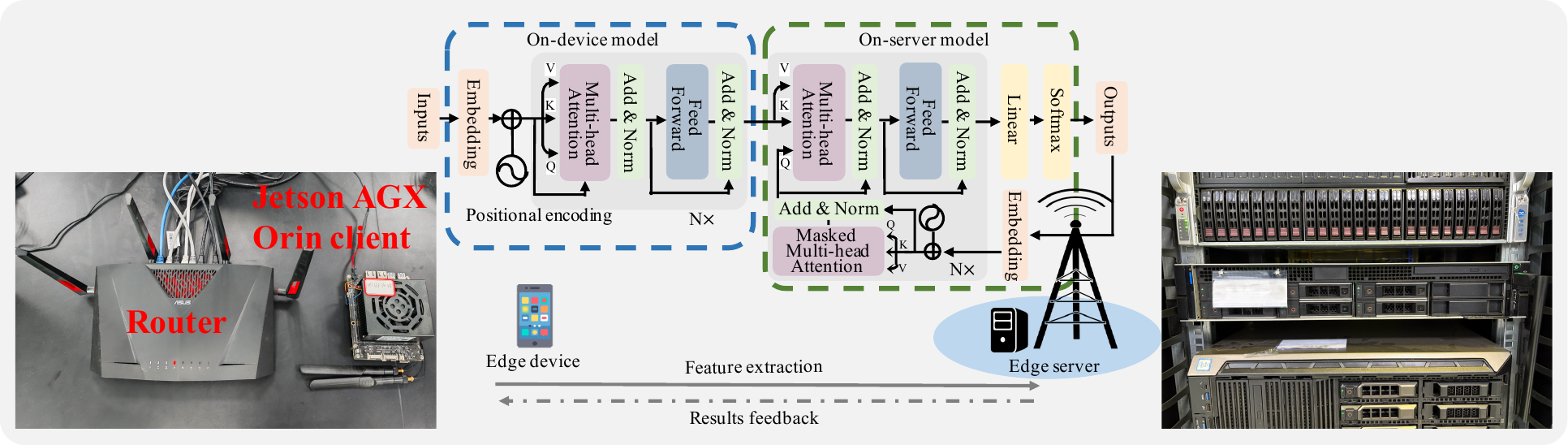}
  \caption{Real-world testbed for pruning-aware LAIM co-inference in sensing-driven edge AI.}
\label{Fig:sensing4AI}
\end{figure*}

\textbf{Viewpoint and geometry adaptation:}  For embodied or spatially distributed sensing, viewpoint control and geometric configuration directly affect observation informativeness. In robotic manipulation, multi-view picking \cite{8793805} develops an active next-best-view policy that selects camera perspectives based on grasp-pose confidence, reducing occlusion uncertainty and improving grasping efficiency. In the aerial domain, \cite{9635893} integrates DRL with semantic segmentation to plan UAV flight paths, enabling UAVs to move toward perceptually informative regions while avoiding visually degraded areas. From a deployment perspective, \cite{10638100} optimizes camera placement by modeling the spatial density of targets of interest and solving a task-oriented coverage problem that prioritizes high-value regions rather than uniform coverage. Beyond physical systems, virtual environments also facilitate the development and transfer of active perception policies. For instance, the Gibson Environment \cite{Xia_2018_CVPR} offers a physics-constrained and photorealistic simulator for learning motion-based perception behaviors that generalize to real-world settings. Extending to open-world reasoning, MP5 \cite{10657187} establishes a multi-modal and goal-conditioned active perception framework in Minecraft, where agents continuously adjust camera poses and navigation trajectories while combining visual observation with language-guided reasoning for long-horizon tasks. Collectively, these studies show that viewpoint and motion control, across robotic, aerial, and virtual environments, form the basis of embodied active perception for edge AI.

\textbf{Multi-agent coordination:}  In multi-robot systems and distributed sensing networks, active perception emphasizes coordinated decision-making to improve collective efficiency and coverage.  {The agentic system in \cite{vyas2025smartadaptiveagentsactive} demonstrates on-device active sensing, where a Jetson-based saccadic agent couples real-time perception and planning through active inference. It autonomously controls the camera field of view to focus on informative regions under tight resource constraints, demonstrating the feasibility of embodied active perception on edge hardware.} In cooperative exploration, Dec-MCTS \cite{doi:10.1177/0278364918755924} develops a decentralized Monte Carlo tree search approach that allows agents to maintain local plans and exchange policies, supporting distributed operation under limited communication resource. Complementarily, \cite{11150751} proposes a distributed multi-robot active sensing framework that combines an extended information consensus filter with control barrier functions to jointly optimize information acquisition and safety, producing collision-free yet informative trajectories. These results highlight how embodied intelligence and inter-agent coordination jointly enable task-oriented perception in large-scale scenarios.

\subsubsection{Resource- and semantic-aware Active Perception}
In resource-constrained camera networks, active perception further requires temporal and energy-aware control, which transforms perception into a context-adaptive function tightly coupled with resource dynamics.
The energy modeling framework in \cite{7517353} quantifies the coupling among sensing rate, processing load, and communication power, providing an analytical basis for frame-rate and duty-cycle adjustment. Building on such insights, \cite{10906058} proposes an adaptive configuration framework that jointly tunes video frame rate, model complexity, and computing resources via a Markov approximation strategy and keyframe selection, maintaining analytics accuracy with reduced system cost. In vehicular edge scenarios, DEVA \cite{LEE2025101467} aggregates in-vehicle devices into a distributed computation pool and dynamically adjusts dashcam frame rates and task pipelines to sustain real-time driving monitoring. 

At a higher level, active perception can move from data acquisition toward semantic-level coupling between perception outcomes and downstream generation or decision making. WiPe-GAI \cite{10472660} unifies wireless perception and generative AI in mobile edge networks, where a sequential multi-scale perception algorithm extracts user motion features from CSI to guide diffusion-based content generation. The framework links sensing, generation, and resource optimization into a closed loop, illustrating how active perception can evolve from low-level parameter control to high-level semantic intelligence at the edge.

\subsection{Case Study and Lesson Learned}

As shown in Fig. \ref{Fig:sensing4AI}, \cite{11301737} validates pruning-aware LAIM co-inference for visual sensing tasks under practical edge deployment constraints. This case study focuses on how a large perception model can be distributedly executed across an edge device and an edge server. The developed testbed consists of an NVIDIA Jetson AGX Orin 64G as the edge device, a Dell PowerEdge R740 server equipped with dual Intel Xeon Gold 6246R CPUs and two NVIDIA RTX 3090 GPUs as the edge server, and an ASUS RT-AC88U router that provides a stable 5 GHz Wi-Fi connection between them. The experiment considers the Kinetics-400 video understanding dataset and adopts VideoMAE as the perception model, which contains 22.03 million parameters and requires 34.29 GFLOPs per inference. 

In this setup, VideoMAE is partitioned into an on-device sub-model and an on-server sub-model. The edge device processes the sensed video input with the front part of the model and uploads the resulting intermediate embeddings to the edge server, which then completes the remaining inference procedure. To make such co-inference feasible on resource-constrained edge hardware, model pruning is introduced to reduce the computation and storage costs of both sub-models, while the pruning ratio and model split point are selected under latency, energy, and hardware-resource constraints. Since the transmit power and computation frequency are governed by hardware-level management mechanisms and are difficult to precisely control in the deployed testbed, the experiment first measures the communication and computation delay/energy of on-device inference, on-server inference, and co-inference under different split points, and then optimizes the pruning ratio based on the measured system parameters. 

The testbed results show that the pruning-aware co-inference design consistently achieves higher inference accuracy than both pure on-device and pure on-server execution. This case study demonstrates that, for sensing-driven edge AI, the practical benefit of co-inference does not only come from offloading computation to a stronger server, but also from matching model compression and model partitioning with the measured communication-computation characteristics of the deployed edge hardware.

{\bf Lesson Learned}: This section reveals that task-oriented edge perception fundamentally depends on closed-loop co-adaptation among sensing, communication, and intelligence, where system configurations are continuously adjusted based on task-level feedback from learning and inference processes.
The ISCC paradigm provides a principled system-level framework to explicitly capture  the intrinsic coupling among sensing quality, transmission reliability, and computational workload under practical QoS constraints. By shifting the optimization objective toward task-centric metrics, ISCC enables coordinated resource allocation across the entire edge AI pipeline. Moreover, active perception transforms sensing from passive data acquisition into selective information gathering by adaptively controlling sensing configurations. Collectively, these insights highlight that effective edge perception is not a static pipeline but a dynamic and feedback-driven process, in which sensing actions and system resources are continually co-adapted to maximize downstream AI task performance.

\begin{table*}[h]
\centering
\caption{Comparison of ISCC and Active Perception Paradigms for Task-oriented Sensing Assisted Edge AI}
\label{tab:task_oriented_sensing_assisted_edge_ai}
\setlength{\tabcolsep}{2pt}
\renewcommand{\arraystretch}{1.4}
\fontsize{7.5}{7.5}\selectfont
{
\begin{tabularx}{\textwidth}{
>{\raggedright\arraybackslash}p{0.8cm}
>{\raggedright\arraybackslash}p{1.9cm}
>{\raggedright\arraybackslash}p{7.1cm}
>{\raggedright\arraybackslash}p{3.6cm}
>{\raggedright\arraybackslash}X}
\toprule
\multicolumn{1}{c}{} 
& \multicolumn{1}{c}{} 
& \multicolumn{1}{c}{\textbf{What is jointly adapted}} 
& \multicolumn{1}{c}{\textbf{Feedback signal/task goal}} 
& \multicolumn{1}{c}{\textbf{Representative scenarios}} \\
\midrule

\textbf{ISCC}
& System-level frameworks
& Sensing rate/quality and sensing-computation resources \cite{10159270}; device association and subchannel allocation \cite{9763441}; unified hardware/protocol co-design \cite{9687468}
& Task utility, resource efficiency, and robust cross-domain coordination
& Edge AI with ISCC orchestration \\

& Edge learning
& Sensing scheduling and dual-layer beamforming \cite{10415206}; sensing, communication, learning co-optimization \cite{9970330}; ISCC design for vertical federated edge learning \cite{cao2025joint}
& Learning loss, convergence behavior, and privacy constraints
& Distributed learning \\

& Edge inference
& Sensing time, transmit power, and quantization bits \cite{10217150}; inference mode selection \cite{10631278}
& Discriminant gain, inference accuracy, communication reliability, and energy constraints
& Multi-agent collaborative  inference \\

& Energy- and delay-aware design
& Sensing rate, offloading ratio, and CPU utilization \cite{11146901}; UAV trajectory, transmit power, and offloading policy \cite{11318338}
& Energy consumption, delay, and environment dynamics
& Intelligent IoT and low-altitude economy \\
\midrule

\textbf{Active perception}
& Parameter adaptation
& Camera parameters \cite{10884674}; RoI scheduling \cite{10202594}; cross-camera workload assignment \cite{10682605}
& Scene complexity, analytics accuracy, and spatio-temporal redundancy
& Edge video analytics and cross-camera surveillance \\

& Viewpoint and geometry adaptation
& Camera pose selection \cite{8793805}; UAV path planning \cite{9635893}; camera placement \cite{10638100}; embodied navigation/pose adjustment \cite{Xia_2018_CVPR,10657187}
& Grasp confidence, semantic informativeness, and perception coverage
& Robotics, UAV perception, surveillance, and embodied/agentic intelligence \\

& Multi-agent coordination
& FoV control \cite{vyas2025smartadaptiveagentsactive}; policy exchange \cite{doi:10.1177/0278364918755924}; robot trajectories \cite{11150751}
& Information gain, safety constraints, and perception coverage
& Embodied/agentic intelligence\\

& Resource- and semantic-aware adaptation
& Frame rate, duty cycle, and SCC resources \cite{7517353}; video frame rate/model complexity/computing resources with keyframe selection \cite{10906058}; dashcam task pipelines \cite{LEE2025101467}; perception-guided semantic generation \cite{10472660}
& System cost, task quality, and generative feedback
& Smart transportation and generative AI \\
\bottomrule
\end{tabularx}
}
\end{table*}

\section{Research Challenges and Open Issues}

Despite rapid progress, edge perception remains far from fully mature, and its large-scale deployment in real-world environments faces a number of fundamental challenges.

\subsubsection{Task-Oriented Performance Modeling}

A central open problem is how to quantitatively characterize how sensing quality impacts downstream edge learning and inference. Sensing decisions, such as power allocation, sampling rate, and measurement resolution, directly shape the quality and diversity of the data that drive model training and prediction. However, this relationship is often non-monotonic and counterintuitive: insufficient sensing causes information loss and degrades learning, while excessive precision may yield diminishing returns, and noisy yet diverse data can sometimes improve generalization through implicit regularization.

The main challenge is to establish task-oriented metrics that connect sensing fidelity and data quality to learning and inference outcomes, including accuracy, convergence speed, and robustness. This requires modeling the end-to-end causal chain, from sensing configuration, to data representation, to model update and computation dynamics, and finally to end-task performance. Promising directions include information-theoretic characterizations, such as rate–distortion–perception trade-offs, as well as data-driven surrogate models that learn quantitative relationships between sensing parameters and task-level metrics through empirical experiments. Another direction is to develop semantic-aware utility functions that evaluate sensing actions by their contributions to decision confidence and task relevance, rather than by raw signal strength or measurement precision.
Such formulations will ultimately enable adaptive sensing control, where sensing parameters are dynamically tuned to maximize task efficiency under real-world QoS constraints. Establishing principled linkages between perception quality and AI task performance is therefore fundamental to transforming edge sensing from passive data acquisition into an active and goal-oriented component of edge AI systems.

\subsubsection{Data Scarcity and Fusion under Resource Constraints}

A fundamental bottleneck in edge perception arises from data scarcity and heterogeneity. Many edge sensing modalities (e.g., radar, LiDAR, and wireless CSI) lack large-scale, well-labeled datasets for model pre-training or fine-tuning. Acquiring labeled and synchronized multi-modal data is expensive and labor-intensive, while real-world measurements are often corrupted by noise, misalignment, and temporal drift caused by mobility, occlusion, and hardware variability. As a result, AI models trained on static or idealized datasets frequently fail in dynamic, unseen, or long-term deployment scenarios. Moreover, large-scale data cleaning and calibration are costly and often impractical, making it difficult to maintain consistent data quality across heterogeneous sensing sources.

Generative (large AI) models provide a potential solution to mitigate data scarcity, but it simultaneously introduces new challenges. Diffusion and foundation models can synthesize missing modalities, generate pseudo-labels, and augment rare events. However, their authenticity, controllability, and bias remain insufficiently understood. For safety-critical applications, such as autonomous driving or robotic surgery, unverified synthetic data may pose severe risks. A key open problem is therefore how to establish governed generative data loops, in which synthetic data are generated, validated, and shared across edge networks with uncertainty estimation and continuous quality monitoring.

 {Even when multi-modal data are available, fusion under edge constraints raises a more fundamental question than how to fuse, namely, when the addition of a modality is truly worthwhile. In edge perception, the value of fusion (VoF) should be assessed by comparing the marginal task-level gain brought by an extra modality or a deeper fusion level against the induced system cost. In general, VoF is high when modalities are strongly complementary and the task operates in ambiguity-prone conditions, but low when the added modality is redundant or the system is already bottlenecked by QoS constraints. Practical deployments are further hindered by modality misalignment and the lack of unified representations for interoperable fusion.}

 {To address these issues, future research should prioritize reliability-guaranteed data generation and uncertainty-aware pseudo-labeling to improve the trustworthiness of synthetic data. It is also important to formalize VoF to guide modality selection, fusion granularity, and execution frequency. In addition, semantic communications and lightweight multi-modal alignment mechanisms are needed to transmit only task-critical features and handle practical impairments in downstream processing.}




\subsubsection{Green and Sustainable Edge Perception}

Energy efficiency is emerging as a central and first-class design objective for intelligent edge perception systems. A prerequisite for sustainable operation is accurate and fine-grained energy modeling that captures the power consumption of the entire perception pipeline, including sensing, data transmission, and AI computation. Existing studies often rely on coarse-grained analytical models that fail to reflect fine-grained energy dynamics across heterogeneous hardware and operating conditions, such as sensing sampling rate, ambient temperature, and server cooling overhead.
Beyond modeling, green edge perception requires system-level energy management. This task is further complicated by device heterogeneity, as energy efficiency varies significantly across sensors, edge devices, and servers. Resource orchestration must therefore jointly consider hardware diversity, task priority, and long-term energy efficiency.

The growing adoption of renewable and energy-harvesting technologies further reshapes the design space of edge perception systems. Modern sensors, edge devices, and servers are increasingly equipped with energy-harvesting capabilities, such as solar, wind, or RF energy harvesting. While these sources create new opportunities for sustainable and carbon-aware operation, they also introduce significant uncertainty in energy availability, prediction, and scheduling, particularly when coordinating intermittently harvested energy with conventional grid-powered supplies.

\begin{table*}[t]
\centering
\caption{Representative Datasets, Testbeds, and Prototypes for Benchmarking Edge Perception}
\label{tab:benchmark_edge_perception}
\setlength{\tabcolsep}{2.2pt}
\renewcommand{\arraystretch}{1.4}
\fontsize{7.5}{7.5}\selectfont
{
\begin{tabularx}{\textwidth}{
>{\raggedright\arraybackslash}p{2.65cm}
>{\raggedright\arraybackslash}p{0.92cm}
>{\raggedright\arraybackslash}p{1.08cm}
>{\raggedright\arraybackslash}p{1.38cm}
>{\raggedright\arraybackslash}p{2.38cm}
>{\raggedright\arraybackslash}p{2.48cm}
>{\raggedright\arraybackslash}p{1.32cm}
>{\raggedright\arraybackslash}X}
\toprule
\multicolumn{1}{c}{\textbf{}}
& \multicolumn{1}{c}{\textbf{Type}}
& \multicolumn{1}{c}{\textbf{Modality}}
& \multicolumn{1}{c}{\textbf{Task}}
& \multicolumn{1}{c}{\textbf{\makecell[c]{Deployment\\setting}}}
& \multicolumn{1}{c}{\textbf{\makecell[c]{Edge\\constraints}}}
& \multicolumn{1}{c}{\textbf{\makecell[c]{Licensing /\\privacy}}}
& \multicolumn{1}{c}{\textbf{\makecell[c]{Main gap to\\practical deployment}}} \\
\midrule

\textbf{DAIR-V2X \cite{9879243}}
& Dataset
& Camera + LiDAR
& Cooperative 3D detection
& Real-road V2X scenes with paired vehicle and infrastructure sensing
& Temporal asynchrony, delay
& Public release, privacy desensitization
& Edge deployment validation under realistic QoS constraints and network dynamics \\

\textbf{WaterScenes \cite{10571852}}
& Dataset
& 4D radar + camera + GPS/IMU
& Detection and segmentation
& Unmanned surface vehicle data under diverse environments
& Adverse environments, sensor failure cases
& Public release
& Edge deployment validation \\

\textbf{Network ISAC testbed \cite{JiComMag2023}}
& Testbed
& 5G NR ISAC
& Cooperative sensing
& Two cooperative ISAC systems with hardware implementation and field trial
& Sensing-communication resource tradeoff
& -
& Benchmarking across modalities, tasks, and deployment conditions \\

\textbf{SenseORAN \cite{10353027}}
& Testbed
& Cellular / O-RAN
& Radar detection
& O-RAN stack with near-RT RIC and YOLO-based xApp
& Fully overlapped LTE/radar signals, near-RT control latency
& Privacy concern on raw I/Q data
& Broader task validation and more reusable benchmarking \\

\textbf{ESP32 WiFi sensing \cite{9217780}}
& Prototype
& WiFi
& Human activity recognition
& ESP32-based standalone or smartphone-attached sensing
& Battery capacity, embedded deployment
& Open-source tool
& Broader modalities and tasks \\

\textbf{WiFi sensing networks \cite{9525003}}
& Prototype
& WiFi
& Soil sensing and precision mapping
& ESP32-based mesh sensing network
& Battery capacity, embedded deployment
& -
& Broader modalities and tasks \\

\textbf{MilliEye \cite{10.1145/3450268.3453532}}
& Prototype
& Radar + camera
& Object detection
& Edge deployment with collected multi-modal dataset
& Lightweight edge inference, limited labeled multi-modal data
& -
& Labeled data scale and benchmark standardization \\

\textbf{LiDAR AIoT roadside prototype \cite{10815971}}
& Prototype
& LiDAR
& Real-time 3D perception
& Edge-cloud-terminal roadside AIoT
& Real-time inference
& -
& Platform openness and cross-vendor reproducibility \\

\textbf{UAV RGB-thermal onboard system \cite{speth2022deep}}
& Prototype + dataset
& RGB + thermal
& Monitoring and person detection
& UAV with onboard NVIDIA AGX Xavier
& Onboard resource capacity
& Publicly released dataset
& Broader edge perception tasks and platforms \\

\textbf{Geryon \cite{10.1145/3550298}}
& Prototype
& Radar + camera
& Object detection
& UAV with nearby edge server
& Edge resources, inference delay
& -
& Broader edge perception tasks and platforms \\

\textbf{Wearable fall-detection device \cite{A2024e28688}}
& Prototype
& Wearable sensing
& Fall detection
& Jetson Nano / Raspberry Pi boards
& Onboard resource capacity and inference latency
& Ethics approval
& Broader validation on population diversity and long-term deployment reliability \\

\textbf{ISCC platform \cite{10631278}}
& Prototype
& Wireless sensing
& Edge AI inference
& ISCC platform with two USRPs
& Edge resources and accuracy-energy-latency tradeoff
& -
& Standardized public benchmarking with reusable data and protocols \\
\bottomrule
\end{tabularx}
}
\end{table*}

To address these challenges, future research should pursue joint model-driven and data-driven energy modeling that combines theoretical analysis with empirical measurement to capture how key parameters, such as sampling rate and hardware utilization, affect real power consumption. Moreover, real-time energy management should integrate energy-availability forecasting with adaptive, energy-aware sensing and computation control, including event-triggered sensing, selective sleep-wake mechanisms, and cooperative energy sharing among nodes. Such predictive control can be further enhanced by learning-based energy schedulers, for example RL or diffusion-based models, that continuously refine energy-consumption models and control policies online. Ultimately, green edge perception calls for a holistic framework that unifies energy modeling, harvesting, and control, enabling perceptual intelligence that is accurate and responsive while remaining self-sustaining and environmentally responsible.

\subsubsection{Trust, Security, and Privacy}

Security and privacy are critical concerns in edge perception. Continuous multi-modal sensing inevitably exposes sensitive information, including facial identities, physiological signals, and spatiotemporal trajectories, which creates persistent risks of privacy leakage and unauthorized profiling. In addition, sensing streams across modalities such as RF, radar, cameras, and LiDAR are susceptible to spoofing and physical adversarial perturbations, potentially corrupting environmental understanding and downstream task execution. At the algorithmic level, distributed learning and inference across heterogeneous edge nodes further introduce sophisticated attack vectors, such as backdoor attacks, gradient inversion, model poisoning, and side-information leakage, which threaten both model integrity and data confidentiality.

Addressing these risks requires a comprehensive security and privacy framework tailored to edge perception. At the data level, life-cycle management mechanisms are needed to regulate the collection, storage, transmission, and processing of multi-modal sensing data. Privacy-preserving preprocessing should be integrated into perception pipelines, for example through feature anonymization and differential-privacy-based perturbation. At the device level, reliable perception further depends on authentication and provenance tracing, such as hardware fingerprints and watermarking, to ensure sensing-source trustworthiness. At the algorithmic level, distributed learning and inference must incorporate privacy protection and adversarial robustness, including secure aggregation, differential privacy, and backdoor detection and mitigation. Collectively, these measures can enable trustworthy edge perception systems that protect sensitive information, resist manipulation, and sustain reliable performance under adversarial and privacy-constrained environments.

\subsubsection{Standardization and Benchmarking}
Standardizing edge perception across vendors, devices, and runtime environments remains a major challenge. Although standardization bodies such as 3GPP, ETSI, and IEEE have begun incorporating wireless sensing into their specifications, multi-modal edge sensors still lack concrete and unified interfaces for downstream AI tasks. In particular, key elements, such as camera intrinsics/extrinsics, radar waveform and antenna configurations, LiDAR calibration parameters, coordinate frames, and point-cloud attributes, are described in vendor-specific and often incompatible formats, which hinders cross-vendor interoperability and multi-modal fusion. Moreover, edge perception pipelines increasingly rely on transmitting compressed features, yet shared definitions of feature tensor layouts, bitstream formats, normalization and quantization parameters, as well as robustness guarantees under packet loss or compression errors, are largely absent. Fragmentation is further exacerbated by the parallel evolution of cellular ISAC (3GPP), WLAN sensing (IEEE 802.11bf), and vision-based sensing ecosystems, which expose incompatible control interfaces and capability descriptions. Differences in operating systems, sensor drivers, AI accelerators, and runtime environments further cause the same perception task or model to behave inconsistently across platforms. Finally, most existing prototypes and testbeds remain limited to single-modality or single-vendor settings, preventing realistic end-to-end evaluation of edge perception under heterogeneous and non-ideal deployment conditions.

To enable large-scale deployment, standardization efforts should prioritize a set of concrete and practical objectives. First, unified metadata and interface specifications should define both raw-sensing and feature-level ``perception I/O'', covering  fields such as timestamps, coordinate frames, and feature schema (tensor layout) so that perception modules can interoperate across heterogeneous devices. Second, lifecycle standards should specify deployment compatibility requirements, such as model identifiers, supported operators, drivers, and runtime versions, to ensure consistent model behavior across hardware and vendors. Finally, the community should establish open, multi-modal, multi-vendor benchmarking testbeds for representative scenarios such as low-altitude aerial platforms, together with unified reporting protocols that jointly measure task utility, latency, bandwidth/compute footprints, and robustness under different scenarios. Moving forward, coordinated efforts among 3GPP, ETSI, IEEE, and relevant industry alliances will be essential to build an interoperable, portable, and rigorously testable ecosystem for edge perception.

\section{Conclusions}
This survey presented a unified, edge-perception-centric review of the closed-loop integration of sensing and edge AI, emphasizing the fundamental limitations of isolated sensing-centric or edge-AI-centric design paradigms for emerging perception-driven services in next-generation wireless networks. From this perspective, we systematically reviewed foundational edge sensing and edge AI techniques, and then examined their intrinsic bidirectional interactions, including edge AI empowered sensing and task-oriented sensing-assisted edge AI. Looking ahead, we identified key research challenges and open issues toward enabling efficient, trustworthy, and scalable edge perception systems, and highlighted promising directions for the development of tightly integrated perception-intelligence architectures toward 6G and beyond.

\bibliographystyle{IEEEtran}
\bibliography{AirCompforFL}

\end{document}